\documentclass[aps, prl,reprint,a4paper,superscriptaddress,amsmath,amsfonts,amssymb, 
]{revtex4-2}
\usepackage{bm}
\usepackage[caption=false]{subfig}
\usepackage{tikz}
\usetikzlibrary{shapes.misc,positioning,shapes}
\usetikzlibrary{patterns}
\usepackage{array, multirow}
\usepackage[colorlinks,linkcolor=blue,citecolor=blue]{hyperref}%
\usepackage{relsize}

\usepackage{times}

\usepackage[normalem]{ulem}

\usepackage[english]{babel}
\usepackage{blindtext}

\newcolumntype{P}[1]{>{\centering\arraybackslash}p{#1}}

\begin{document}
\title{No-Collapse Accurate Quantum Feedback Control via Conditional State Tomography}

\author{Sangkha Borah}
\email{sangkha.borah@mpl.mpg.de}
\affiliation{Max Planck Institute for the Science of Light, Staudtstra{\ss}e 2, 91058 Erlangen, Germany}
\affiliation{Department of Physics, Friedrich-Alexander-Universit\"at Erlangen-N\"urnberg, Staudtstra{\ss}e 7, 91058 Erlangen, Germany}
\affiliation{Okinawa Institute of Science and Technology, Okinawa 904-0495, Japan}
\author{Bijita Sarma}
\email{bijita.sarma@fau.de}
\affiliation{Department of Physics, Friedrich-Alexander-Universit\"at Erlangen-N\"urnberg, Staudtstra{\ss}e 7, 91058 Erlangen, Germany}
\affiliation{Okinawa Institute of Science and Technology, Okinawa 904-0495, Japan}


\begin{abstract}
The effectiveness of measurement-based feedback control (MBFC) protocols is hampered by the presence of measurement noise, which affects the ability to accurately infer the underlying dynamics of a quantum system from noisy continuous measurement records to determine an accurate control strategy. To circumvent such limitations, this work explores a real-time stochastic state estimation approach that enables noise-free monitoring of the conditional dynamics including the full density matrix of the quantum system using noisy measurement records within a single quantum trajectory -- a method we name as `conditional state tomography'. This, in turn, enables the development of precise MBFC strategies that lead to effective control of quantum systems by essentially mitigating the constraints imposed by measurement noise and has potential applications in various feedback quantum control scenarios. This approach is particularly useful for reinforcement-learning (RL)-based control, where the RL-agent can be trained with arbitrary conditional averages of observables, and/or the full density matrix as input (observation), to quickly and accurately learn control strategies.
\end{abstract}

\maketitle
Future advancements in quantum technologies will hinge on the ability to effectively manipulate quantum systems by controlling their states through reliable protocols and feedback strategies~\cite{wiseman_milburn_book, Zhang2017Mar, Jacobs_book, Doherty2000Jun}. Broadly speaking, pure control strategies entail using open-loop pulse-based controls for quantum circuits, and such problems have been successfully tackled using conventional optimal control tools like gradient-ascent pulse engineering (GRAPE)~\cite{deFouquieres2011Oct, Morzhin2019Oct, Koch2022Dec, Sarma2022Nov}. These methods are fundamentally based on a differentiable model of quantum dynamics that cannot be extended to feedback-based controls~\cite{deFouquieres2011Oct, Propson2022Jan}. For controls employing continuous measurement, non-trivial strategies need to be identified based on conditional dynamics. These measurement-based feedback control (MBFC) techniques are considered pivotal for achieving real-time quantum control in laboratory experiments~\cite{Martin2020Oct, Kuang2021Aug, Rossi2018Nov, Vijay2012Oct, Tebbenjohanns2021Jul, Wilson2015Aug,  Livingston2022Apr, Magrini2021Jul, JimenezMartinez2018Jan}. Reinforcement learning (RL) has recently been proven as a powerful new ansatz for such control tasks, which, in the quantum domain, was first demonstrated  for quantum error correction~\cite{Marquardt2018_drl_qec} and optimization of quantum phase transition in 2018~\cite{Bukov2018Sep}. Following these initial studies, we have recently witnessed its applications in different sets of non-intuitive problems, including applications in quantum control~\cite{Sarma2022Nov,Ueda2020_quantum_cartpole_drl,Niu2019Apr,Zhang2019Oct,Borah2021_double_well}, state transfer~\cite{Prati2019_drl_stirap,Paparelle2020}, quantum state preparation and engineering~\cite{Porotti2022Jun,Wang2019_drl_state_prepare,Wrachtrup2020_drl_state_engineering,Haug2020_drl_state_preparation_classification}, and quantum error correction~\cite{Nautrup2019Dec}. Very recently, the use of RL controls for real laboratory experiments of a quantum system has become a reality~\cite{Reuer2022Oct, Sivak2022Nov}.

At a fundamental level, the MBFC approaches based on continuous measurements suffer from the limitations of two primary sources. First, such approaches often fail to control the dynamics beyond a specific limit set by the signal-to-noise ratio of the intrinsic and unavoidable measurement-induced noise to the measured quantity. The level of noise increases as $1/{\sqrt{\kappa  \delta t}}$, where $\kappa$ denotes the measurement rate, and $\delta t$ is the measurement time interval, which given the fact that $\delta t$ is related directly to the variance of the noise distribution (in the Wiener noise model) and $\delta t \ll 1$, the actual measured signal can be well hidden in the sea of random noise~\cite{Borah2021_double_well}. This makes it practically impossible for MBFC to find suitable control strategies for the system to achieve the desired dynamics. Second, the continuous measurement process naturally leads to the so-called measurement backaction, which makes the MBFC schemes highly non-intuitive and nontrivial in general~\cite{Hacohen-Gourgy2020Jan, Zhang2017, Marquardt2018_drl_qec, Borah2021_double_well, Porotti2022Jun}.  

In this Letter, we research in this direction and propose an efficient MBFC protocol that can precisely control the dynamics of a quantum system of interest based on noisy, continuous, and real-time measurement data. This is made possible by developing a measurement-based stochastic estimator that can extract the real-time state of the measured system noiselessly and without collapse, thereby controlling the system dynamics in any desired way. Unlike the usual method of state estimation with continuous measurement using thousands of trajectories from that many copies of the quantum system, this method estimates the conditional state of the system from single trajectory, a method we term as `conditional state tomography'. We demonstrate the efficiency of the scheme by applying it to control the dynamics of linear and nonlinear quantum systems where the applied feedback is state-based or conditional. We also show the usefulness of the scheme for cases where control laws can be derived based on conditional moments (assuming perfect extraction of the measured signal from the noisy data, which is typically not possible in realistic experiments), which we illustrate with an example of preparing symmetric and antisymmetric entangled states of two qubits. Moreover, our scheme is adaptable for real-time feedback with RL controllers, allowing optimal and efficient training and control.

\begin{figure}[t]
    \centering   
    \includegraphics[width=1.0\linewidth]{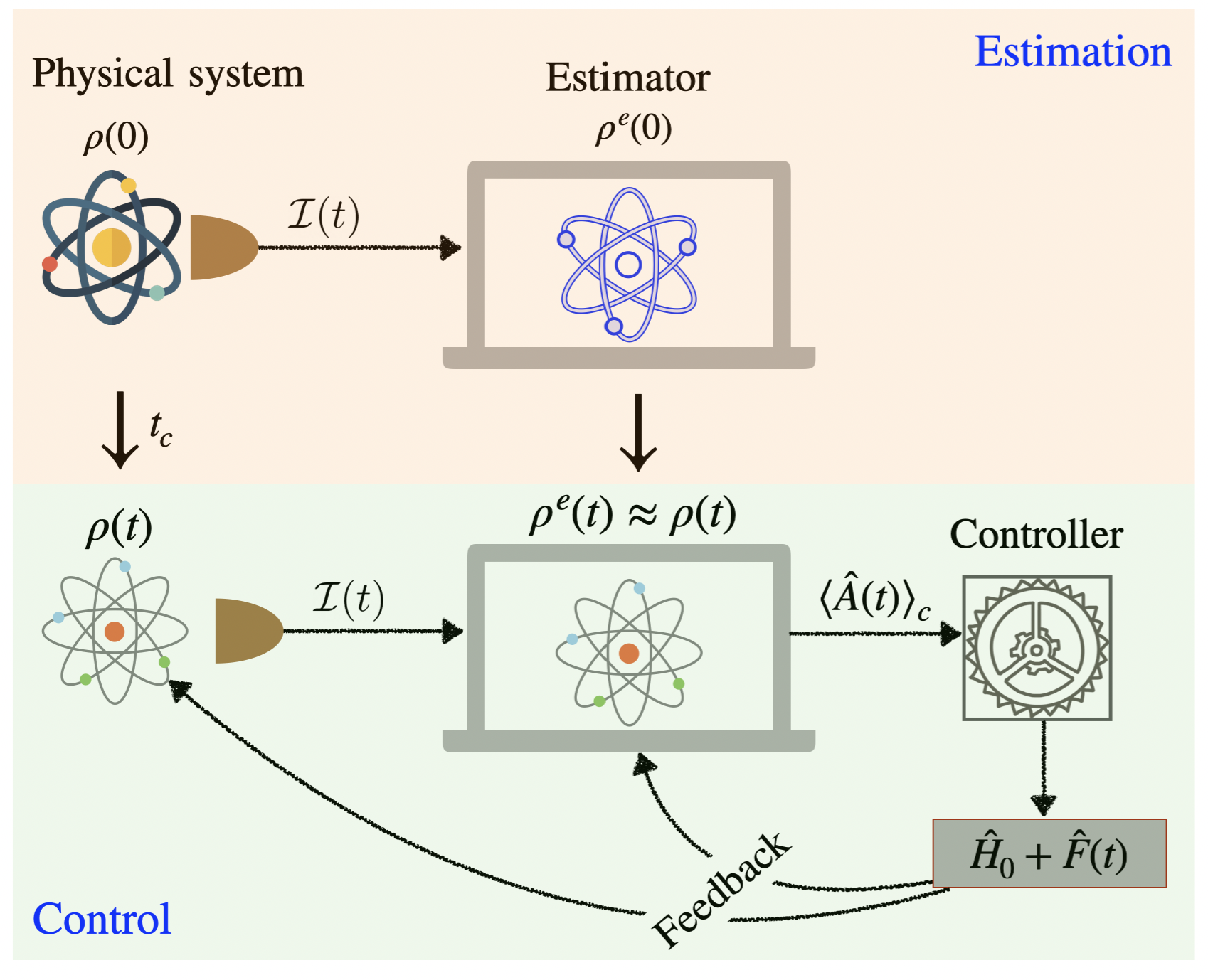}
    \caption{The schematic of the proposed protocol. (Top) The estimation stage: a physical quantum system (left) described by Hamiltonian $\hat{H}_0$ is continuously monitored to probe the observable $\hat{A}$, and the noisy measurement outcomes are fed to the estimator (right) -a \textit{simulator} based on the mathematical model of the real physical system on a classical processor, eg.~a FPGA (Field Programmable Gate Array). The state of the physical system (estimator) at time $t$ is described by $\rho(t)$ ($\rho^e(t)$) which becomes equal at $t \ge t_c$. (Bottom) The control stage: a controller is used to apply accurate feedback $\hat{F}(t)$ to both the physical as well as the estimator systems as a function of the estimated noiseless conditional signal $\langle\hat{A}(t)\rangle_c$ obtained through the estimator.}
    \label{fig:scheme}
\end{figure}

The protocol is shown schematically in Fig.~\ref{fig:scheme}. It consists of two operation steps, the estimation stage and the control stage. In the estimation stage, the to be controlled quantum system (shown on the left), with an unknown initial state (given by the density matrix $\rho(0)$) is measured using a (weak) continuous measurement approach. The noisy current streams from the measurement are then used to construct a stochastic estimator (shown on the right), which is a computational model of the measured quantum system, with the same Hamiltonian but with any random initial quantum state $\rho^e(0)$. The estimator can track the dynamics of the measured quantum system in real-time after a while, as the conditional state of the estimator converges to that of the physical quantum system. In the control stage of operation, a controller is developed to mediate between the real system and the estimator by applying feedback on the systems based on the conditional dynamics of the latter while continuing to control the systems through the real-time measured data of the physical quantum system. 

We first describe the theory behind the measurement-based stochastic estimator and the feedback control method. Suppose the laboratory quantum system (top left in Fig.~\ref{fig:scheme}), with Hamiltonian $\hat{H_0}$, is being measured continuously with a weak probe for the measurement operator $\hat{A}$ (suitably scaled to make it dimensionless). Such a continuous measurement process leads to conditional stochastic dynamics of the system density matrix in time $\rho_c(t)$ and is described by the so-called quantum stochastic master equation (SME),
\begin{align}
\nonumber
\frac{d\rho_c(t)}{dt} =  - i [\hat{H_0}, ~\rho_c(t)]+ \kappa \mathcal{D}[\hat{A}] \rho_c(t) 
\\+ \sqrt{\kappa \eta} \mathcal{H}[\hat{A}] \rho_c(t) d\xi(t).
\label{eq:SME}
\end{align}
Here, $\kappa$ is the measurement rate (the rate at which information is extracted from the detector), $\eta$ is the measurement efficiency of the detector and $d\xi(t)$ represents an instantaneous random Wiener noise increment (white noise model with zero mean and variance $\sqrt{dt}$, where $dt$ is the time interval between successive measurements). $\mathcal{D}[\hat{A}]$ and $\mathcal{H}[\hat{A}]$ are the superoperators describing respectively the backaction and diffusion terms in the SME~\cite{wiseman_milburn_book, Jacobs_book}, see Supplemental Material for more details. 
Probing the system with a weakly coupled meter that, in effect, has a broad probability distribution of the quantum state leads to noisy measurement records given by,
\begin{align}
{\cal I}(t) = \langle \hat A(t) \rangle_c+ \frac{1}{\sqrt{4\kappa\eta}} {d\xi(t)}.
\label{eq:measurement_current_first_occurrence}
\end{align}
The first term on the right-hand side of the above equation denotes the conditional mean of the measurement operator (the signal) and the second term represents the contribution of the measurement noise, which depends on $\eta$ and $\kappa$.

The estimator is a model quantum system with the same Hamiltonian $\hat{H_0}$, as depicted in Fig.~\ref{fig:scheme} (top right), which is initialized in any arbitrary quantum state $\rho^e (0)$, and is driven by the noisy measurement current of the real laboratory quantum system, ${\cal I}(t)$~(Eq.~\ref{eq:measurement_current_first_occurrence}). The dynamics of the estimator is described by the modified SME~\cite{wiseman_milburn_book, Diosi2006_coupled_ito, Zhang2017}, 
\begin{align}
\nonumber
d\rho^e_c(t) = & - i [\hat H_0, ~\rho^e_c(t)] dt + \kappa \mathcal{D}[\hat{A}] \rho^e_c(t) dt \\
+& {2\kappa \eta}\left[
{\cal I}(t) - \langle A(t)\rangle_c^e  \right] \mathcal{H}[\hat{A}] \rho^e_c(t) dt,
\label{eq:mbe_scheme_main_eq}
\end{align}
where $\rho^e_{c} (t)$ denotes the conditional density matrix of the estimator independent of the real system, and $ \langle \hat{A}(t) \rangle_c^e = \mathrm{Tr}[\rho^e \hat{A}]$ is the conditional mean calculated for the estimator at time $t$. In essence, the estimator dynamics is driven by the noisy real-time measurement currents from the meter and the conditional means of the estimator itself. It can be shown that the overlap between the states $\rho(t)$ and $\rho^e(t)$ following Eqs.~\ref{eq:SME} and \ref{eq:mbe_scheme_main_eq} monotonically increases until it reaches unity: $
    \delta \mathrm{Tr}[\rho\rho^e](t) \sim   \mathrm{Tr}\big[\sqrt{\rho} ( \hat{A} 
    + \langle \hat{A} \rangle) ~\rho^e \big( \hat{A} 
    + \langle \hat{A}  
    \rangle\big)
    \times \sqrt{\rho} \big] \delta t. $
Thus, provided the estimator gets sufficient amount of measurement data, the convergence of its dynamic state to that of the physical quantum system, i.e., $\rho_e(t) \sim \rho(t)$ can always be guaranteed, except for the cases where $[\hat{H}_0, \hat{A}] = 0$. In case of the latter, the observable $\hat{A}$ is a constant of motion, and the continuous measurement of it does not provide any information about the state of the system, causing $\rho^e(t)$ to remain in one of the eigenstates of $\hat{H_0}$. It is possible to ensure convergence regardless of the values of $\eta$ and $\kappa$, although the time it takes to reach convergence, $t_f$, will be longer if $\eta$ and $\kappa$ are lower (see the Supplemental Material for an example of a qubit to illustrate the protocol). Once this estimation stage is complete, the second stage of the MBFC scheme, namely the control stage, is initiated, as shown in Fig.~\ref{fig:scheme}(bottom). 

\begin{figure}[t]
    \centering
\includegraphics[width=1.0\linewidth, trim={0.0cm 0.0cm 0 0},clip]{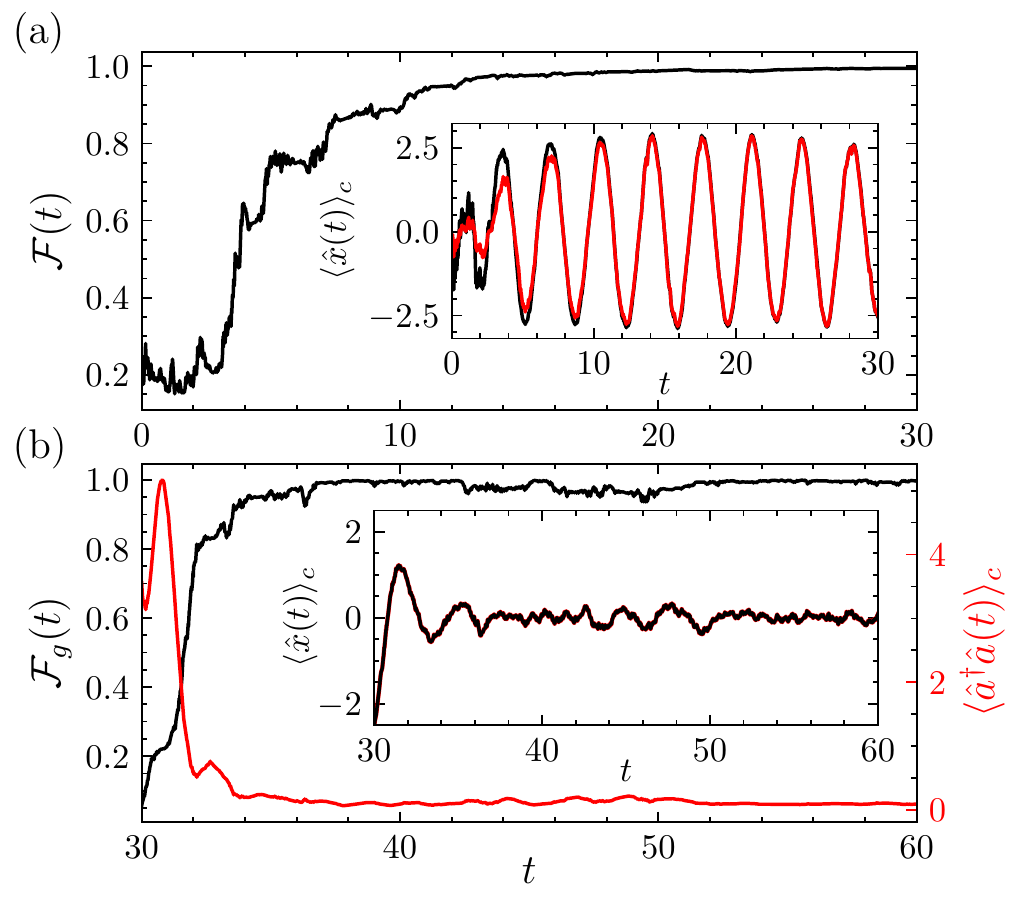}
  \caption{Control of a linear quantum harmonic oscillator using the protocol. (a) In the estimation phase, the fidelity $\mathcal{F}(t)$ between the physical system and the estimator steadily converges. The inset displays the conditional means of the observable $\hat{x}$. (b) Subsequently, a state-based controller is applied, swiftly guiding the particle's motion around the center $\langle \hat{x}(t)\rangle_c = 0$. The instantaneous fidelity $\mathcal{F}_g(t)$ (depicted in black) quantifies the closeness between the physical system/estimator state and the target state, i.e., the ground state of the oscillator. The conditional mean population $\langle \hat{a}^\dagger \hat{a} (t)\rangle_c$ in the oscillator is shown in red. }
    \label{fig:QHO}
\end{figure}

We first apply the scheme for dynamic feedback cooling of a linear quantum harmonic oscillator and demonstrate how it becomes possible to employ accurate state-based feedback control to achieve this. The Hamiltonian of the linear quantum harmonic oscillator is given by $\hat H_0 = {\hat p^2}/{2m} + m \omega^2 \hat x^2/2,$
where $\hat x$ and $\hat p$ are the position and momentum operators, respectively, $m$ is the mass of the oscillator, and $\omega$ denotes the frequency of oscillation. Consider making a measurement of the position operator such that $\hat{A} = \hat{x}$. 
In Fig.~\ref{fig:QHO}(a), the instantaneous fidelity between the states of the real system and the estimator, $\mathcal{F}(t)$ is shown during the estimation stage of the control protocol. As shown in terms of the monotonically improved fidelity, the estimator starts mimicking the dynamics of the measured quantum system; also shown in the inset of the Fig.~\ref{fig:QHO}(a), where the evolution of the conditional means of $\hat{x}$ for the measured system and estimator are compared. After the estimation stage is completed, which is typically smaller than $\kappa^{-1}$, the control stage is initialised. 
We now use a state-based control strategy given by $\hat H(t)  = \hat H_0 - \langle \hat x (t) \rangle_c \hat p$, where $\langle \hat x (t) \rangle_c $ denotes the conditional mean of $\hat x$ at time $t$. 
Such a feedback represents a damping control scheme, where the controller applies feedback based on the conditional mean of the position operator to effectively reduce the momentum as it approaches $\langle\hat{x}(t)\rangle_c \to 0$.
The feedback is applied to both the measured system and the estimator based on the noise-free conditional mean of the position extracted by the estimator. The results are shown in Fig.~\ref{fig:QHO}(b), where it is found that the proposed control protocol leads to fast and accurate dynamic cooling of the quantum harmonic oscillator. The inset of Fig.~\ref{fig:QHO}(b) shows how the control protocol could keep the quantum state at a dynamical minimum to any length of time, which is crucial. 

\begin{figure}[t]
    \centering
\includegraphics[width=1.0\linewidth]{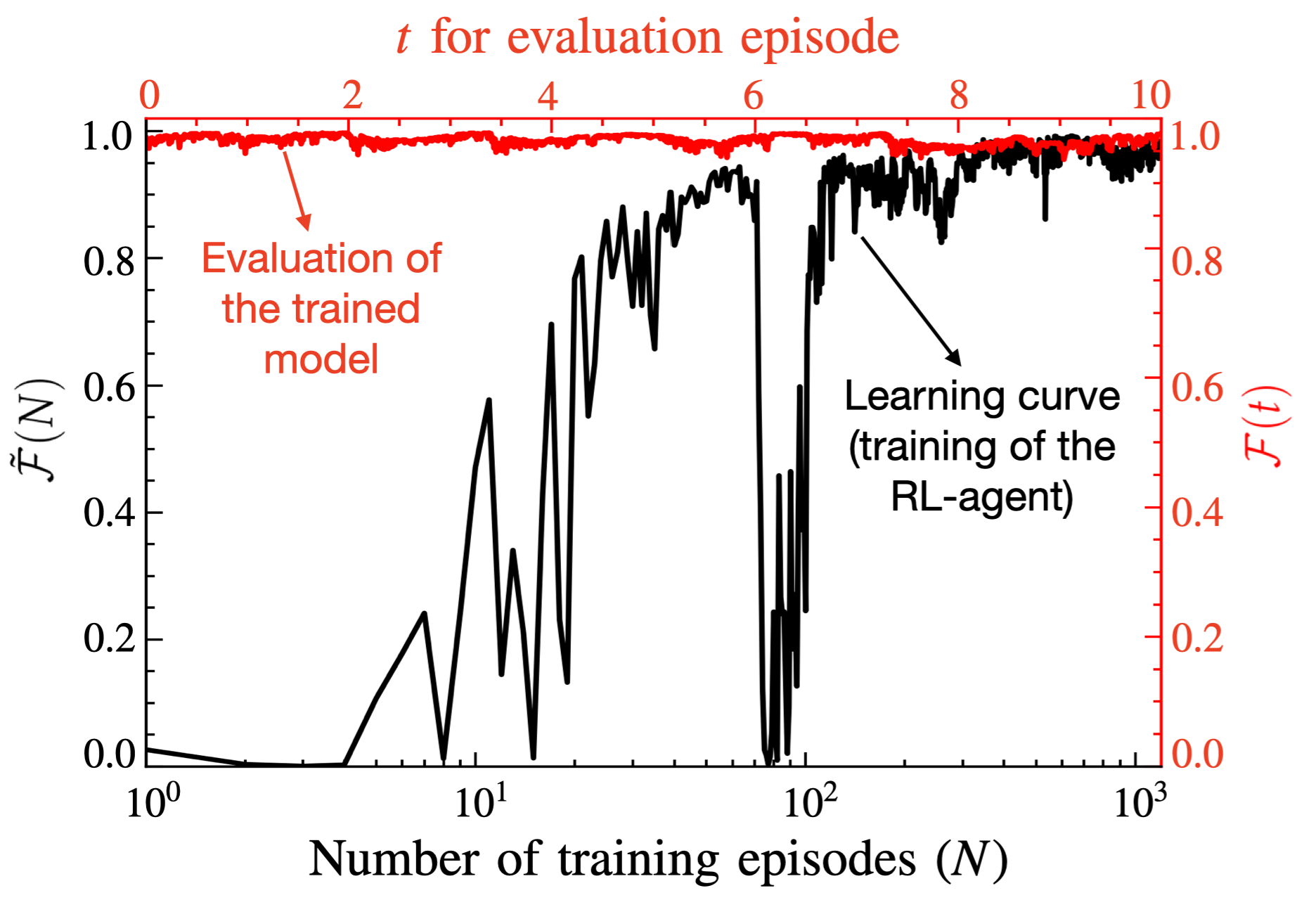}
  \caption{The protocol is applied to control a particle's motion in a nonlinear quartic potential to cool it to its dynamic ground state using RL-based control. The training process is shown in black colored line as the average fidelity over each episode $N$ with respect to the target state (ground state), $\bar{\mathcal{F}}(N)$, which is maximized through training. Note that the sudden drop at $N\sim 100$ is due to the exploration of the RL-agent. The performance of the trained agent is shown in red line. 
  }
    \label{fig:QQO}
\end{figure}

Next, we consider a nonlinear quartic potential with the unperturbed Hamiltonian given by $\hat H_0 = {\hat p^2}/{2m} + \lambda \hat x^4,$ where we have chosen $m=1/\pi$ and $\lambda = \pi/25$ with proper dimensions. We apply artificial control {\it{viz}}.~RL~\cite{Sutton_drl_book, ppo_paper, trpo_paper}, to devise proper feedback strategies in this case. It is noteworthy that with the designed stochastic estimator, it is now possible to apply the full density matrix as well as the means and moments of the operators for choosing any accurate feedback scheme. Therefore, the scheme allows using accurate conditional means of observables as the input $s_t$ (observation) to the RL-agent; for example, here we use $s_t = \{ \langle \hat{x}\rangle, \langle\hat{p}\rangle\}$. Another advantage of the estimator control is that state fidelities are now realizable, which are usually pervasive in real experimental measurements. Therefore, given that we have access to the fidelity $\mathcal{F}(t)$ of the estimator, it can be used as a simple and efficient reward function that needs to be maximized by the RL-agent in the training process. 
The agent is first trained with a given initial state, which, due to the generalizability of the trained model, permits to be used for controlling the system started with other (random) initial states. The learning curve as the mean fidelity $\bar{\mathcal{F}}(N)$ over each training episode $N$ is shown in black in Fig.~\ref{fig:QQO}. Using conditional means for training the RL-agent makes learning quicker and more accurate. The evaluated episodic fidelity variation $\mathcal F(t)$ is shown in red colored line in Fig.~\ref{fig:QQO} in the bi-axial plot's second scale, demonstrating accurate feedback control by the trained RL-model. 
\begin{figure}[t]
    \centering
    \includegraphics[width=1.0\linewidth]{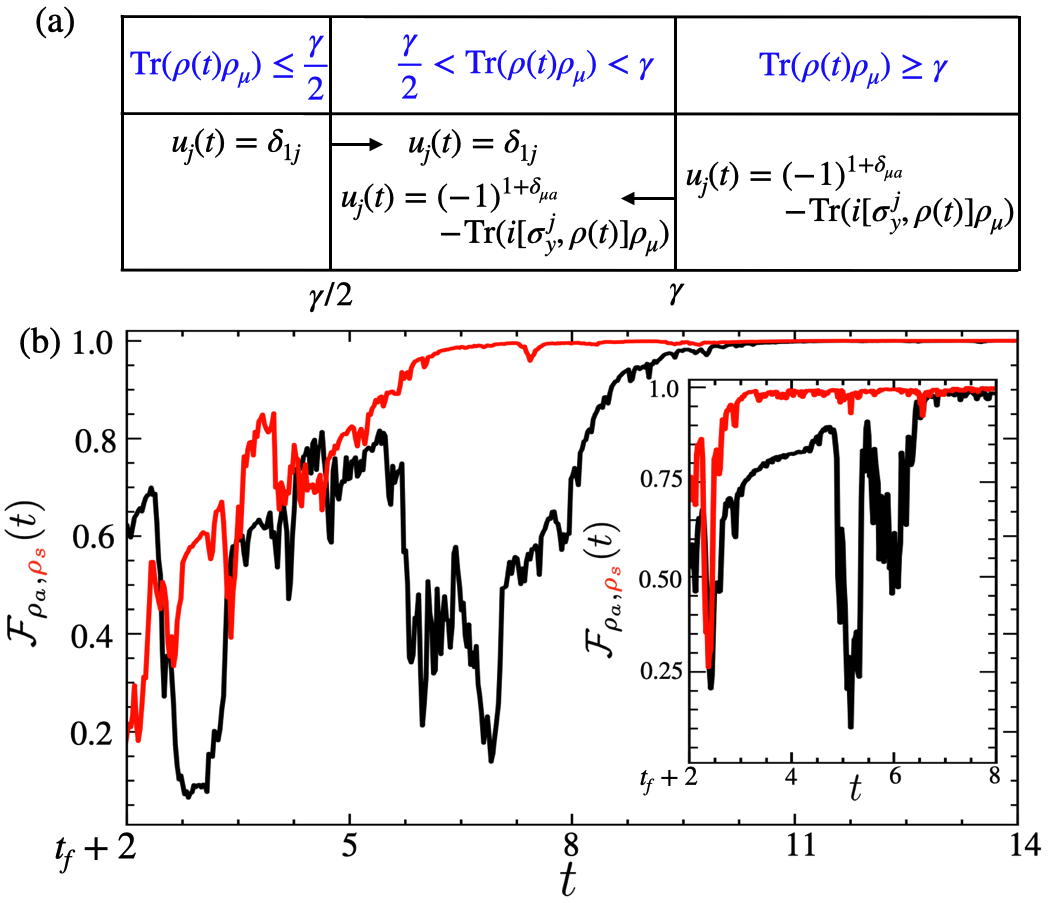}
  \caption{Demonstration of the proposed MBFC protocol for the preparation of symmetric, $\rho_s$, and antisymmetric, $\rho_a$, entangled states between two qubits as an example for when it is possible to derive control laws based on conditional moments within stochastic dynamics. Control laws $u_1$ and $u_2$ are selected depending on the conditional value of ${\rho(t) \rho_\mu}$, where $\mu \in \{s, a\}$ (symmetric and antisymmetric) are in the three regimes, conveniently demonstrated in (a), and the arrows represent the direction of the entrance boundary of $\rho(t)$ to the middle section. $\gamma$ is the damping parameter, the measurement rate $\kappa$ is assumed to be $0.1$, and the efficiency $\eta = 0.5$ for this simulation.  After the estimation stage (not shown), these control laws are applied on conditional mean data (density matrices to compute instantaneous fidelity), which leads to convergence to the target states ($\rho_a$: black and $\rho_s$: red), shown in (b). In the absence of such laws, RL can be used - the performance is shown in the inset of figure (b) with similar color settings. }
\label{fig:entanglement}
\end{figure}

Besides, it is often possible to derive control laws for systems undergoing continuous measurement based on the conditional means of observables (without the noise component). Although such control laws would not have much value in realistic situations due to the unavailability of accurate noiseless signal, we now, show in the following that in such context too, our proposed scheme would be useful. To illustrate it, we consider the preparation of symmetric ($\rho_s$) and antisymmetric ($\rho_a$) entangled states of two qubits, where the states are given by
$
\rho_{(s/a)} = \frac{1}{2}(\psi_{\uparrow\downarrow} \pm \psi_{\downarrow\uparrow})(\psi_{\uparrow\downarrow} \pm \psi_{\downarrow\uparrow})^*.
$
Here, $\psi_{\uparrow \downarrow} = (\uparrow) \otimes (\downarrow)$ and  $\psi_{\downarrow \uparrow } = (\downarrow) \otimes (\uparrow)$ are the tensor product states of the individual qubit states in the ground and excited states. The quantum filtering equation under feedback with control variables $u_1(t)$ and $u_2(t)$ is given by,
\begin{align}
\nonumber
d\rho(t) = & 
- i u_1(t) [\sigma_y^{(1)}, \rho(t)] dt 
- i u_2(t) [\sigma_y^{(2)}, \rho(t)] dt \\ \nonumber
- & \frac{1}{2} [F_z, [F_z, \rho(t)]] dt 
+ \sqrt{\eta} \big\{F_z\rho(t) 
+  \rho(t)F_z \\
-& 2~{\rm Tr}[F_z \rho(t)]\rho(t)\big\} dW_t,
\label{eq:control_laws_u1_u2}
\end{align}
where $dW_t$ is the Winner noise increment at time $t$. $\sigma_{g}^{i}$, $g \in \{x, y, z\}$ and $i=\{1, 2\}$ are tensored Pauli operators for qubit $i$ and $F_z = \sigma_z^1 + \sigma_z^2$~\cite{Mirrahimi2007_qubit_control}. The control laws dictate non-intuitive choices of the control parameters $u_1(t)$ and $u_2(t)$ provided the real-time conditional fidelity between the current and the target states, $\rho_s$ and $\rho_a$ could be accurately extracted via conditional tomography of the quantum states, which is often a difficult task if not impossible. These are discussed in the Supplementary Material and conveniently represented in Fig.~\ref{fig:entanglement}(a). Using these control laws with the MBFC scheme makes it possible to evaluate the controls $u_1(t)$ and $u_2(t)$ in real time which leads to a guaranteed preparation of the states $\rho_a$ and $\rho_s$, shown in black and red lines, respectively, in Fig.~\ref{fig:entanglement}(b). It becomes also possible to use RL for control similar to the case shown for a quartic oscillator above, in which case one can use the full density matrix for training along with conditional means, and the performance is shown in the inset of the figure. Compared to the control laws, the RL controller can help the system reach its target state in a shorter time scale.

Finally, we will mention possible shortcomings of the proposed scheme. First, the protocol leans towards a model-based approach, aiming to maximize controlled output accuracy based on a highly precise physical model, and therefore one should care about potential model bias. To remove model bias, one can integrate model learning techniques such as Hamiltonian learning beforehand~\cite{Gebhart2023Mar}. It is also possible to use machine learning techniques such as Bayesian estimation~\cite{Nolan2021Dec} and RL~\cite{Xiao2022Jan, Xu2019Oct} for estimating model parameters. Second, when dealing with real-time feedback control problems, it is likely to have potential delays between the measurement and feedback operations. The estimator, being a simulator in a classical processor, needs finite time for simulation that can add to this delay event, especially for large systems. While the estimation stage of the protocol can be streamlined by completing it in a single pass by providing all previous measurement results at once to the estimator; for the control stage, it would be advantageous to provide the estimator and the controller with frequent measurement results, to discover finer controllability, tailored to the system's complexity. In such cases, RL-based methods can be especially effective~\cite{Reuer2022Oct, Sivak2022Nov}. 

In conclusion, even when employing sophisticated noise filtering techniques such as Linear Quadratic Regulator (LQR), Linear Quadratic Gaussian (LQG), and Kalman filters in standard MBFC experiments, extracting the exact signal from the noisy measurement results remains a formidable task~\cite{JimenezMartinez2018Jan}. Consequently, conventional feedback strategies fall short of achieving accurate control. The proposed protocol circumvents this by estimating accurate conditional state tomography, thereby enabling precise quantum feedback control within the realm of continuous measurement. Furthermore, this protocol integrates seamlessly with RL-based control methods, enabling efficient training and control.

\section*{Acknowledgements} 
The authors thank Gerard Milburn, Jason Twamley and Michael Kewming for useful discussions. 

%

\newpage
\widetext

\def\theequation{S\arabic{equation}}
\renewcommand{\thepage}{S\arabic{page}} 
\renewcommand{\thesection}{S\arabic{section}}  
\renewcommand{\thetable}{S\arabic{table}}  
\renewcommand{\thefigure}{S\arabic{figure}}
\setcounter{figure}{0}
\setcounter{table}{0}
\setcounter{section}{0}
\setcounter{subsection}{0}
\setcounter{page}{1}
\begin{center}
	\textbf{\large Supplemental Material}
\end{center}

\section{S1. Continuous measurement theory}
Quantum continuous measurement approach is useful for following the process of changes in the state of a quantum system being measured as one gradually obtains information about it. It is particularly important for situations where a system needs to be monitored in continuous time and active feedback is applied in real time. The dynamics of a quantum system undergoing continuous measurement is described by the so-called stochastic master equation (SME)~\cite{Jacobs_book, wiseman_milburn_book}. 

One approach to obtaining the SME is to correlate it to classical continuous measurement, as outlined in ~\cite{Jacobs_book}. In classical continuous measurement theory, the likelihood of obtaining measurement results $y$ given the true value $x_{\text{true}}$ of a system measured in terms of the continuous variable $x$ is represented by a Gaussian distribution. The conditional probability distribution is represented as
\begin{equation}
    P(y|x) = \frac{1}{\sqrt{2\pi\sigma}} \exp\left[-\frac{(y-x)^2}{2\sigma} \right],
\end{equation}  
where, $\sigma$ is the noise variance, and $\xi$ represents unbiased Gaussian noise. The measurement outcomes are given by $y = x_{\text{true}} + \xi$. The sum of measurement results can be expressed as an integral over increments:
\begin{equation} 
y = \int_0^T x \, dt + \beta \int_0^T dW. \end{equation}
Here, $dW$ represents Wiener noise, obeying $dW^2 = dt$ and $\sum_n (dW_n)^2 = T$. Differential equations involving $dW$ are handled using stochastic calculus. The stochastic differential equation (SDE) for the probability distribution $P(x)$ based on measurements is given by,
\begin{equation}
  dP(x) = \frac{(x - \langle x \rangle)}{\beta^2} \left[ dy - \langle x \rangle \, dt \right],
\end{equation}
where the measurement outcomes are given by,
\begin{equation}
  dy = \langle x \rangle \, dt + \beta \, dW. 
\end{equation} 
This is the Kushner-Stratonovich equation, which characterizes classical continuous measurement theory.

Now, to formulate the quantum continuous measurement theory, we choose a quantum observable $X$ for which the measurement result $dy$ is given by, \begin{equation}
    dy = \langle X \rangle dt + \frac{1}{\sqrt{8k}} dW,
\end{equation}
where we have redefined $k = 1/(8\beta^2)$. The second term on the right-hand side is called the measurement noise or shot noise. The likelihood function $P(dy|x)$ has the form
\begin{equation}
    P(dy|x) = \sqrt{\frac{4k}{\pi dt}} \exp\left[-4k \frac{(dy - xdt)^2}{dt} \right].
\end{equation}
From this, the set of quantum operators $A(dy)$ can be derived as follows,
\begin{align}
    A(dy) = \int_\infty^\infty \sqrt{P(dy|x)} |x\rangle \langle x| dx 
    =  \left(\frac{4k}{\pi dt}\right)^{1/4} \exp\left[-2k \frac{(dy - Xdt)^2}{dt} \right],
\end{align}
where we assume that the measurement operator $X$ has a continuous spectrum of eigenvalues $x$ such that $X|x\rangle = x|x\rangle$ and $\langle x|x^\prime\rangle = \delta(x-x^\prime)$, i.e., the measurement operators are Gaussian weighted sum of projectors onto the eigenstates of $X$. The probability distribution of the measurement results with initial state $|\psi\rangle = \int \psi(x) |x\rangle dx$ can be determined as follows, 
\begin{align}
    P(dy) = \mathrm{Tr}\left[ A(dy)^\dagger A(dy) |\psi\rangle \langle \psi| \right]
    =  \sqrt{\frac{4k}{\pi dt}} \exp\left[-\frac{4k(dy - \langle X \rangle dt)^2}{dt}\right],
\end{align}
where we have replaced $|\psi(x)|^2 \to \delta (x - \langle X \rangle)$, using the fact that the Gaussian in the integral is much wider than $\psi(x)$ and $\langle dy \rangle = \langle X \rangle dt$, so it must be centered at $\langle X \rangle$.  With this, we can now easily derive the SDE for $|\psi\rangle$ as follows,
\begin{align}
    \hat \psi(t + dt) \rangle = A(dy) |\psi(t)\rangle 
    \propto  \left(1 - [k X^2 - 4 k X \langle X \rangle ] dt + \sqrt{2k} X dW\right) |\psi(t)\rangle,
\end{align}
where $|\hat\psi(t + dt)\rangle$ represents the normalized wave function after a measurement. The normalization leads to, 
\begin{align}
    d|\psi\rangle ~=~  \{ - k (X - \langle X \rangle)^2 dt 
    + \sqrt{2 k} (X - \langle X \rangle)dW \}  |\psi(t)\rangle,
    \label{eq:SSE}
\end{align}
which is known as the stochastic Schr\"odiner equation (SSE). Unlike the Schr\"odinger equation, the SSE is essentially nonlinear, since the term containing $\langle X\rangle$ is contained in the term containing $dW$. From this, the stochastic master equation (SME) for the evolution of the density matrix $\rho$ can be easily derived by taking $\rho(t + dt) = (|\psi\rangle + d|\psi\rangle) (\langle \psi | + d\langle \psi|)$ as,
\begin{align}
\nonumber
    d\rho = - \frac{i}{\hbar}[H, \rho] dt -k [X, [X, \rho]] dt 
    +& \sqrt{2k} (X\rho + \rho X - 2\langle X \rangle \rho ) dW,
\label{eq:SME}
\end{align}
where the first term on the right side represents the 
coherent evolution due to the system Hamiltonian $H$. This is the primary equation for the quantum analogue of the classical Kushner-Stratonovich equation. Defining the following superoperator and replacing $k \to \kappa/2$,
\begin{align}
    \mathcal{D}[A]\rho = & A\rho A^\dagger - \frac{1}{2}(A^\dagger A \rho + \rho A^\dagger A),\\
    \mathcal{H}[A]\rho = & A\rho + \rho A^\dagger - \rho ~\mathrm{Tr}[A\rho + \rho A^\dagger],
\end{align}
the SME can be written as, 
\begin{align}
\nonumber
d\rho_c(t) =  - i [H, \rho_c(t)] dt + +\kappa \mathcal{D}[A] \rho_c(t) dt 
+ \sqrt{\kappa} \mathcal{H}[A] \rho_c(t) dW(t).
\end{align}
where $\rho_c$ denotes the conditional density matrix of the system described by the Hamiltonian $H$. $A$ is the dimensionless measurement operator. We refer to continuous measurement records as,
\begin{align}
dQ(t) = \langle A (t) \rangle_c dt + \frac{1}{\sqrt{4\kappa }} {dW(t)}.
\end{align}

\section{S2. Measurement-Based Feedback control}
\begin{figure}[!hbt]
    \centering
    \includegraphics[width=0.7\linewidth]{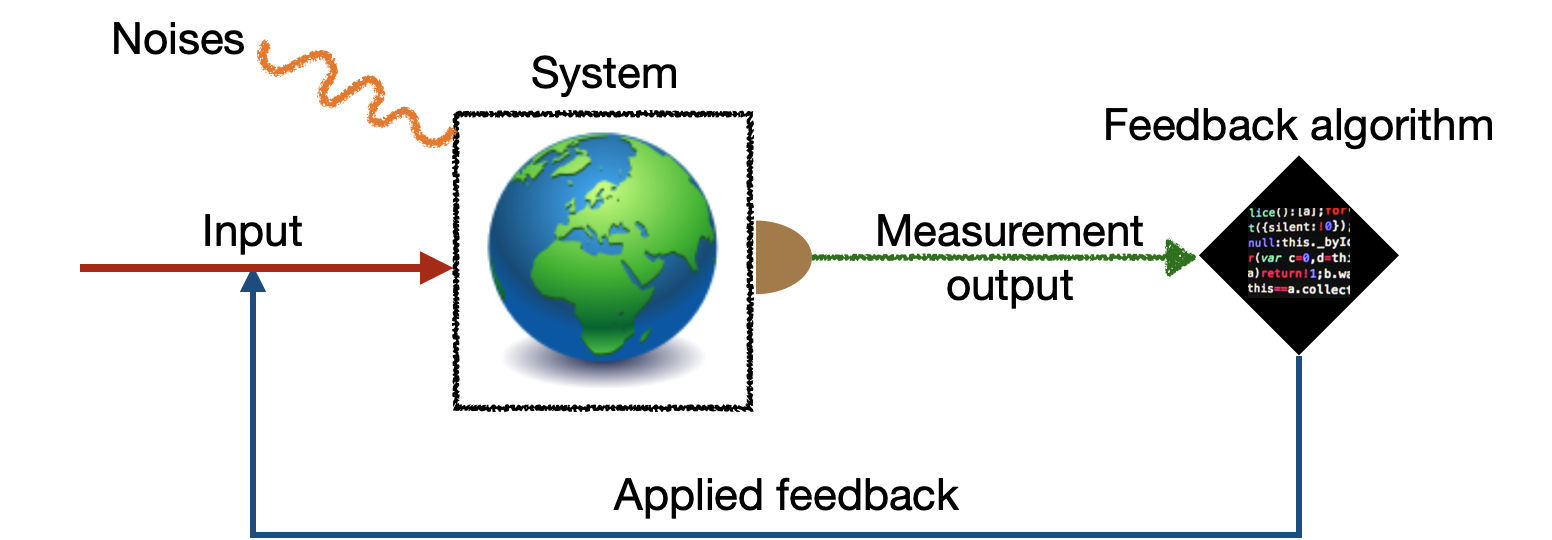}
    \caption{The typical workflow of measurement-based feedback control (MBFC) is shown. A system (left) is measured continuously, and the noisy measurement results are used by a feedback algorithm to change the system's inputs in a desired way and control the system's dynamics.}
    \label{fig:MBFC}
\end{figure}

In a typical Measurement-Based Feedback control (MBFC) scheme, a dynamic system is continuously monitored by a probe that outputs noisy readings in real time, which are processed by a real-time feedback algorithm to modify the input to the system to achieve a desired dynamic or target state over time. This is shown schematically in Fig.~\ref{fig:MBFC}.
The feedback algorithm is usually called a controller and is designed to be a computer or configurable logic block that interprets signals from noisy data and, based on that, applies feedback controls according to some predefined, hard-coded rules. This may be a classical or a quantum system. In the latter case, the signal is also a function of the backaction of the measurement process, in addition to the presence of various noises originating from decoherence effects of the environment. We are interested here in quantum mechanical control by active feedback. Such active feedback (also known as closed-loop feedback control) is based on continuous monitoring of the quantum system with weak measurements, as described above. When the feedback controller applies feedback $F(t)$ to the dynamic quantum system at time $t$, the SME becomes
\begin{align}
\nonumber
d\rho_c(t) =  - i [H, \rho_c(t)] dt +\kappa \mathcal{D}[A] \rho_c(t) dt 
+ \sqrt{\kappa } \mathcal{H}[A] \rho_c(t) dW(t) - i [F(t), \rho_c(t)] dt. 
\end{align} 
Here, the feedback $F(t)$ can generally be a function of all previous measured values $dQ$; in this case, the feedback is essentially non-Markovian. In the case of Markovian feedback, $F(t)$ is determined only by the instantaneous value of the measurement result $dQ(t)$. This is the essence of the so-called Wiseman-Milburn feedback based on measurements~\cite{wiseman_milburn_book}. In the case of Bayesian feedback control, $F(t)$ is a function of the conditional density matrix $\rho_c$ instead of $dQ$. For example, $F(t)$ can be determined by the conditional mean of an operator $X$, which is not readily available for most complex quantum mechanical systems~\cite{Jacobs_book, wiseman_milburn_book}. The distinctive feature of the present work is that we have shown how it may be possible to access noiseless conditional dynamics using noisy continuous measurement techniques that leads to accurate MBFC.

\section{S3. Reinforcement learning controller}
Reinforcement learning (RL)~\cite{Sutton_drl_book} is a type of machine learning (ML)~\cite{Goodfellow_ml_book} that is used to make a series of decisions on complex and potentially uncertain problems. It is trained through a game-like process in which a player learns to play a game from scratch in order to win and accumulate as many points as possible in a shorter time. The ML model in the case of RL is commonly referred to as the RL-agent. The world outside the RL-agent is called the RL-environment, which in principle includes everything, but must include the part of the world defined by the particular problem over which the RL-agent wants to have control. Similarly, the RL-agent also gathers knowledge about the RL-environment by collecting incentives called rewards while trying to change its dynamics (its game) by imposing some stimuli called actions. The RL-agent's goal is to collect as many rewards as possible over a given period of time, called an episode. Thus, the RL-agent essentially learns by trial and error, much like humans and animals, and because of this similarity, RL is considered the primary technique for general machine intelligence. RL is used in numerous engineering applications, including robot navigation, artificial intelligence in games, real-time decision making, skill acquisition, and learning tasks. 

\begin{figure}[!hbt]
\centering
\includegraphics[width=0.7\linewidth]{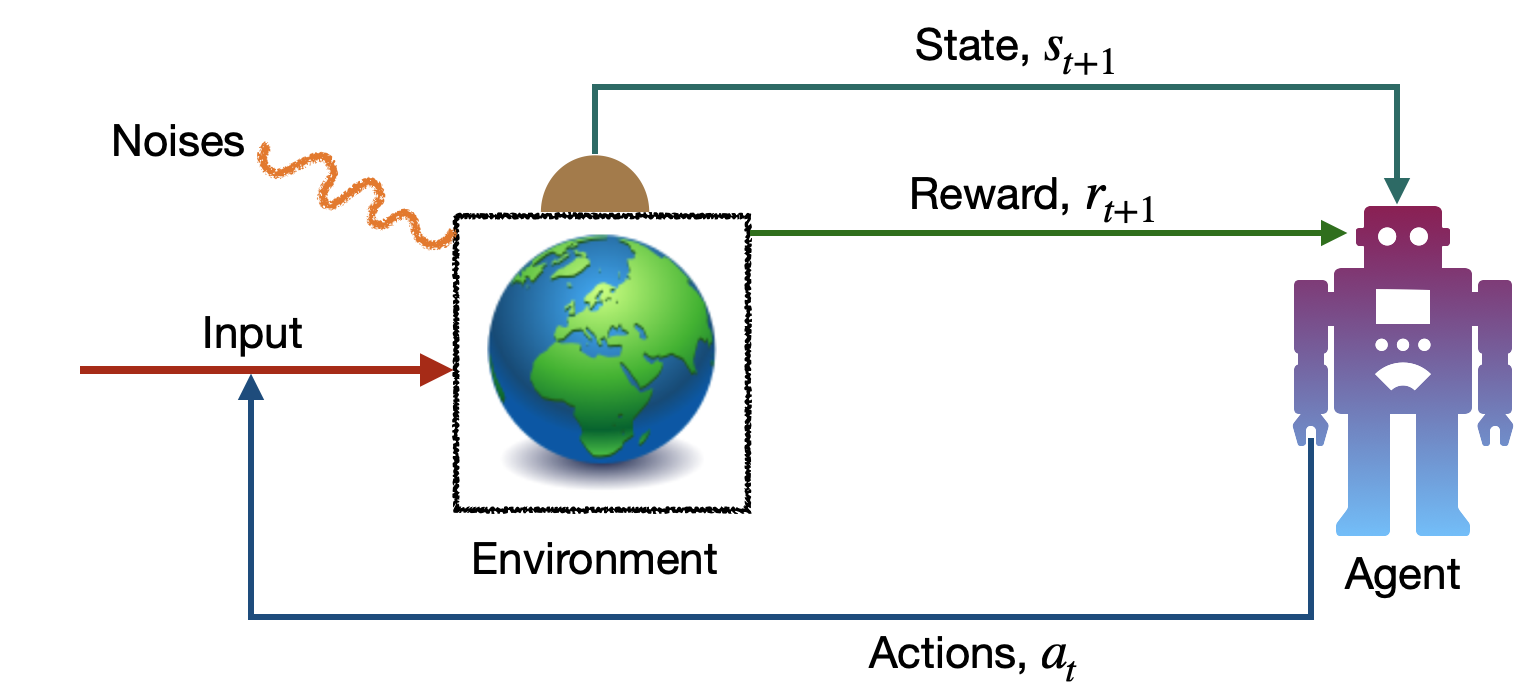}
\caption{Workflow of RL in comparison to MBFC shown in Fig.~\ref{fig:MBFC}. In RL terminology, the controller on the right is called the agent, the system to be controlled by the agent is called the environment, the outputs are called the states and the feedback is called the actions. Learning of the rules of the control (policy) is done via maximization of the net reward it gets over some time called an episode. The reward should any metric of the changes occurred in the environment because of the application of an action and should be a quantity accessible in real experimental conditions.  }
\label{fig:rl_workflow}
\end{figure}

The typical workflow of RL is very similar to the MBFC shown in Fig. \ref{fig:MBFC} above. In the case of RL, the feedback algorithm (controller) is the RL-agent and the system is the RL-environment. The outputs that the RL-agent uses to train itself and learn about the RL-environment and controller are called states or observations. Unlike the controller in MBFC, the RL-agent does not need to be pre-programmed, but is able to learn the rules (policy) of the controls itself based on the observations it receives in real-time from the RL-environment. To do this, the RL-agent must receive another signal, the reward, which, as mentioned above, is maximized over each episode to obtain the optimal strategy. With these modifications, the workflow of MBFC with RL control takes the form shown in Fig.~\ref{fig:rl_workflow}. The definitions of the various RL-terminologies along with brief definitions and notes are given in Table~\ref{tab:RL_termniology}.

RL in conjunction with ANN is referred to as Deep Reinforcement Learning - DRL for short, which has revolutionized the field of RL as a cutting-edge technology in recent years. Although the primary task of RL is to learn to control the environment through actions, there are numerous algorithms to accomplish this. Choosing the right algorithm depends on the nature of the problem we need to solve and the complexity of the actions and observations. In most cases, it turns out that the choice is not trivial and there is always an optimal choice of algorithms and various so-called hyper-parameters that work optimally for a given problem. Therefore, it is overwhelming for novices to learn the procedure in detail, and a chronological understanding is desirable. When applied to problems in quantum physics, the complexity arises not only from the choice of the type of algorithms, but also from the inherent complexity resulting from the uncertainty associated with quantum dynamics. This requires a basic understanding of the different types of RL algorithms. 

The RL problem can be formulated mathematically using a Markov decision process (MDP) that represents the dynamics of the environment during the time that the agent takes actions  in a given state. To this end, the MDP is equipped with a transition function (or transition model) that can predict the state of the environment as a function of the current state and the action taken. The MDP is also equipped with a reward function that can be obtained as a function of the current state of the environment and possibly the action and the next state predicted from it. Thus, the dynamics of the MDP are described by the transition and reward functions, collectively referred to as the "model." In most contexts, the MDP has a model, which means that the transition and/or reward functions are not available.

We have used Proximal Policy Optimization (PPO)~\cite{ppo_paper} which is a reinforcement learning algorithm that is used to optimize the policy of an agent in an environment that is designed to be both simple to implement and effective in practice. PPO is a variant of the popular actor-critic algorithm, which separates the policy (the actor) from the value function (the critic). The PPO algorithm can be broken down into the following steps:
\begin{enumerate}
\item Collect a batch of samples by interacting with the environment using the current policy. These samples consist of a sequence of state-action-reward tuples $(s_t, a_t, r_t)$.
\item Estimate the value function $V(s_t)$ for each state in the batch using a neural network. The value function can be estimated using the Bellman equation, which gives the optimal value function $V^*(s)$ for each state,
\begin{equation}
    V(s) = \max_{a} \mathbb{E}[R(s,a) + \gamma V(s')],
\end{equation}
where $s'$ is the next state, and the expectation is taken over the distribution of next states given the current state and action. The value function can be estimated by iteratively updating the estimates using the Bellman equation, this process is known as dynamic programming~\cite{Sutton_drl_book}. One popular method to estimate the value function is using the temporal-difference learning algorithm, which is a type of online, model-free method for estimating the value function.
\item Estimate the advantage function $A(s_t, a_t)$ for each state-action pair in the batch. The advantage function is an estimate of the difference between the expected return and the value function,
\begin{equation}
A(s_t, a_t) = \mathbb{E}[\sum_{k=0}^{\infty} \gamma^k r_{t+k}] - V(s_t).
\end{equation}
\item Use the samples to update the policy network, which is a neural network that maps states to a probability distribution over actions. PPO modifies the actor objective function by using a `clip' function to ensure that the updated policy is not too far from the previous one. The PPO objective function is,
$$L_{\text{PPO}} = \min(r_t(\theta), \text{clip}(r_t(\theta), 1-\epsilon, 1+\epsilon))A(s_t, a_t),$$
where $r_t(\theta) = \frac{\pi_{\theta}(a_t|s_t)}{\pi_{\theta_\text{old}}(a_t|s_t)}$ is the ratio of the new policy to the old policy, $\pi_{\theta}(a_t|s_t)$ is the probability of taking action $a_t$ in state $s_t$ under the current policy, $\pi_{\theta_\text{old}}(a_t|s_t)$ is the probability of taking action $a_t$ in state $s_t$ under the previous policy, $\text{clip}(r_t(\theta), 1-\epsilon, 1+\epsilon)$ is a function that clips the ratio of the new policy to the old policy to the range $[1-\epsilon, 1+\epsilon]$.
\end{enumerate}

The PPO algorithm is an example of trust region policy optimization algorithm~\cite{trpo_paper} that ensures that the updated policy is not too far from the previous one. This makes the optimization process more stable and prevents the agent from over-fitting to the current policy. For that it modifies the objective function of the actor to ensure that the updated policy is not too far from the previous one. The use of the clip function also helps to reduce the variance of the gradient estimates, which can lead to more stable and efficient training.

One of the key advantages of PPO is that it is relatively simple to implement compared to other state-of-the-art algorithms. PPO does not require the use of complex off-policy methods or value function approximations. Another advantage of PPO is that it is a sample-efficient algorithm. It allows the agent to learn from a relatively small number of samples, which makes it well suited to applications where data collection is expensive or time consuming. PPO also has a good performance on high-dimensional and continuous action spaces. It has proven to be a useful algorithm for problems involving quantum systems.

\begin{table*}[!hbt]
\caption{A list of RL terminologies and their definitions along with small notes for each. }
\centering
\begin{tabular}{p{4.0cm}|p{13.5cm}}
\hline
RL Terminology &  Definition \& brief notes\\
\hline
Agent & It is the controller and essentially the brain of the RL, gradually learning about the hidden features of the system being controlled (the environment) in a trail-and-error process. At each point in time, it receives an observation $s$ from the environment, which is mapped by the it to specific feedbacks (actions) $a$ to be applied to the environment.  \\
Environment & It is the world in which the agent lives and interacts  with, in which it learns and decides what actions to apply to it. Anything that the agent cannot arbitrarily change can be considered part of the environment. It is essentially the representation of the problem to be solved and can be based on the real world (e.g., a robot walking in a field), simulations (e.g., board games such as Atari), or hybrids (e.g., self-driving car where the training can be done on simulated models before working in real environments).  \\
State and Observation, $s$ & State represents the complete information about the environment as output at each time step of the RL workflow. Theoretically, the observation represents the complete or partial description of the state. However, in the RL literature, both terms, states and observations, are used as synonyms. In practice, state can be anything that might be useful for the agent to decide the next action wisely, and is usually represented by a real-valued vector, matrix, or higher-order tensor. The set of all valid observations in a given environment is called the observation space, which can be discrete or continuous in nature. \\
Actions, $a$ & These are the feedbacks that the agent applies to the environment based on the current observation, $s$. The actions change the environment and these changes are expected to happen in a desired way. The set of all valid actions in a given environment forms a space known as the action space, which can be discrete or continuous in nature. \\
Reward, $r$ & It is a scalar number, positive or negative, received by the agent from the environment along with the observations, which is a direct measure of whether the action performed by the agent was useful or not in achieving the goal. The reward at each time $t$ is determined by the current observation, the action, and the subsequent observation of the reward function: $r_t = {\cal R}(s_t, a_t, s_{t+1})$, and often simplified as $r_t = R(s_t, a_t)$.  \\
Episodes and Trajectories , $\tau$ & A trajectory represents a sequence of observations and actions $\tau = (s_0, a_0, s_1, s_1, \dots a_{\cal T}, s_{\cal T})$ over a time $\cal T$, called an episode. The agent's task is to collect as much total reward as possible during each episode. \\
Discounted return & The sum of rewards received by the agent due to its actions within a trajectory $\tau$ is called the rate of return $R({\tau}) = \sum_{t=0}^{\cal T} r_t$ or, more precisely, the finite horizon undiscounted rate of return, where all returns, now and later, are treated equally. In practice, the so-called infinite-horizon discounted return $R({\tau}) = \sum_{t=0}^{\cal \infty} \gamma^t r_t$, where $\gamma \in (0, 1)$ is called the discount factor. This gives preference to the reward received now over that to be received later, and makes it mathematically simpler to treat in equations.    \\ 
Policy, $\pi$ & It is the rule or control strategy learned by the agent to apply actions to the environment to achieve something desired. It is essentially represented as a deterministic or stochastic mapping from states to actions. In a stochastic strategy, the action at time $t$ is given by the conditional probability distribution of state $s_t$: $a_t \sim \pi (\cdot |s_t)$. \\ 
Function approximator & It represents the encoding of a function from training examples. Standard approximators are decision trees, neural networks and nearest neighbor methods. The task of the RL is to optimize the so-called policy parameters $\theta$ represented by the function approximator (weight and biases of the neural network) to optimize the policy $\pi_\theta(\cdot|s)$. \\
$V$-value function, $V^\pi(s)$ & It is denoted as the value of a state $s$ and represents the expected return if we start in state $s$ and act according to policy $\pi$ forever thereafter. It is computed as $V^\pi(s) = \underset{\tau \sim \pi}{\mathbb E}[R(\tau)|s_0 = s]$, where $\underset{\tau \sim \pi}{\mathbb E}$ represents the average discounted return starting at $s_0 = s$ and following the policy in each trajectory over multiple trajectories. The optimal value function $V$ results from following the optimal policy, for which it is given by $V^*(s) = \underset{\pi} {\max}\underset{\tau \sim \pi}{\mathbb E}[R(\tau)|s_0 = s]$.  The optimal action can then be calculated as $a^*(s) = \mathrm{arg}\underset{a}{\max} Q^*(s, a)$.\\
$Q$-value function, $Q^\pi(s, a)$ & It is denoted as the value of a state-action pair $(s, a)$ and represents the expected return if we start with the state-action pair $(s, a)$ and act according to policy $\pi$ forever thereafter. It is computed as $Q^\pi(s, a) = \underset{\tau \sim \pi}{\mathbb E}[R(\tau)|s_0 = s, a_0=a]$, where $\underset{\tau \sim \pi}{\mathbb E}$ represents the average discounted return of starting at $s_0 = s$, taking the action $a_0 = a$, and following the policy in each trajectory over multiple trajectories. The optimal $Q$-value function results from following the optimal policy, for which it is given by $Q^*(s, a) = \underset{\pi}{\max} \underset{\tau \sim \pi}{\mathbb E}[R(\tau)|s_0 = s, a_0=a]$. The optimal action can then be calculated as $a^*(s) = \mathrm{arg}\underset{\pi}{\max} Q^*(s, a)$. The main connections between the $V$-value and $Q$-value functions are $V^{\pi}(s) = \underset{a\sim \pi}{\mathbb E} {Q^{\pi}(s,a)}$ and $V^*(s) = \underset{a}{\max} Q^* (s,a)$.\\
Advantage function, $A^\pi(s, a)$ & It is computed as $A^{\pi}(s,a) = Q^{\pi}(s,a) - V^{\pi}(s)$ and represents how much better it is to take a given action $a$ in state $s$ versus a randomly chosen action $\pi(\cdot |s)$. It is important for formulating RL methods for policy optimization. \\
\hline
\end{tabular}
\label{tab:RL_termniology}
\end{table*}

For the results presented in this work, we have demonstrated the use of RL for the non-linear quartic potential control and in the case of preparation of entangled states.  In this protocol, the RL can be trained as a state-aware model of the estimator quantum state or any mean value of interest. For the case of the quartic oscillator, we consider the conditional moments of a few observables as the input to the RL agent,
\begin{equation}
s_t = \{ \langle \hat{x}\rangle, \langle\hat{p}\rangle, \langle \hat{x}^2\rangle, \langle\hat{p}^2\rangle\},
\end{equation}
where $\hat x (\hat p)$ are the position (momentum) quadrature of the oscillator. These, along with the real-time fidelity are provided by the estimator in real-time, the latter is used as the reward signal that gets maximized through training the RL-agent by adjusting the control parameters, 
\begin{equation}
    R(t) = \mathcal{F}(\rho(t), \psi_{\rm ground}).
\end{equation}
The actions, $\lambda(t)$ of the RL (the controls) are real scalar values such that the feedback added to the Hamiltonian is $H_{\rm f} = \lambda(t) \hat p$. 

For the case of preparation of an entangled state based on a control law, the task of the RL is to estimate to controls $u_1$ and $u_2$, as discussed in S5. For this case, the RL uses the conditional density matrix obtained from the estimator as input(observation) to correctly learn the controls in order to generate the target states, using the fidelity as the reward signal. 

For the RL implementation, we used PPO with continuous controls with the following hyperparameters.

\begin{table}[!hbt]
\caption{PPO Hyperparameters}
\begin{tabular}{ l l }
\hline \\
 Policy network size &  $256\times 256$ \\
 Value network size &  $256\times 256$\\
 Batch size & 64 \\ 
 Learning rate  & $10^{-4}$ \\  
 Network update step interval & 2048 \\
 \hline
\end{tabular}
\end{table}

\section{S4. Convergence of fidelity for a driven qubit}

\begin{figure}[!hbt]
    \centering
    \includegraphics[width=0.5\linewidth]{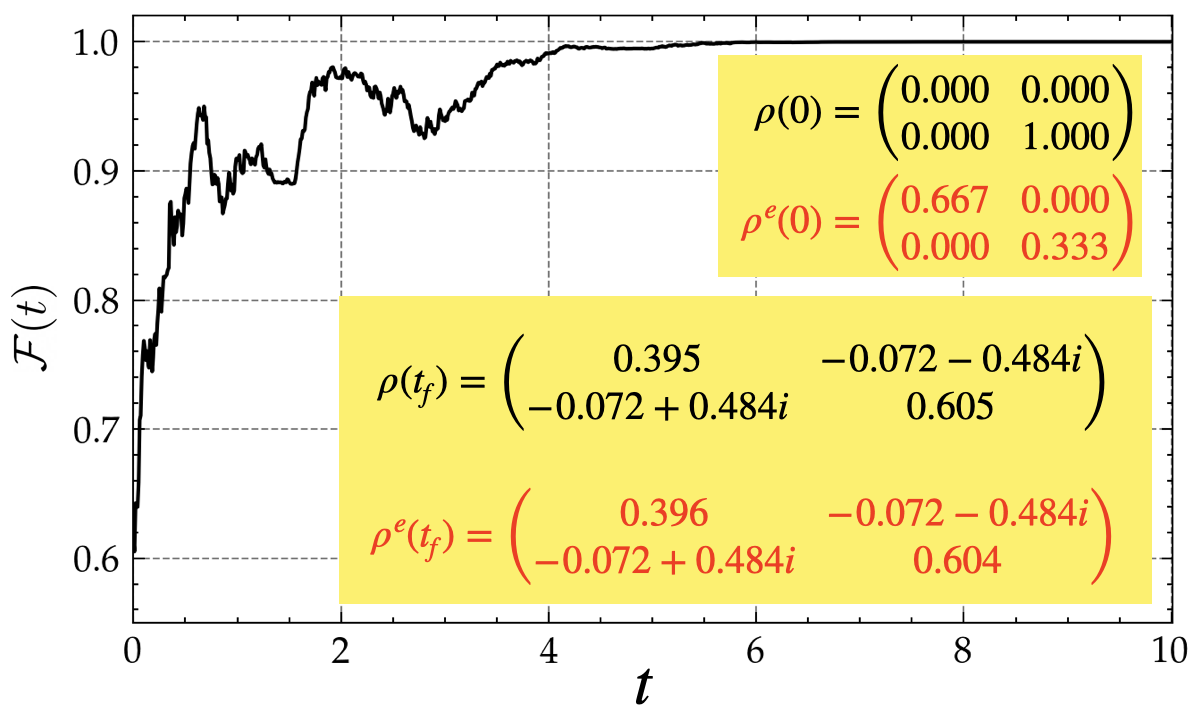}
    \caption{We demonstrate the convergence of the state of the estimator to the real quantum system state {for the toy model of a qubit}. In the insets, we show the initial and final states of the real and the estimator for this particular example. The initial fidelity of the real and estimator states is $\mathcal F(0) \sim 0.6$, which gradually improves until it reaches $\mathcal F(t_f) \approx 1$. This represents the estimation phase of the MBFC protocol shown in Fig.~1(a) in the main text. The parameters considered are $\varepsilon = 0.1, ~ \delta = 1.0, ~ \kappa = 1.0, ~ \eta = 1.0$. }
    \label{fig:estimator_show_example1}
\end{figure}

\begin{figure}[!hbt]
    \centering
    \includegraphics[width=0.5\linewidth]{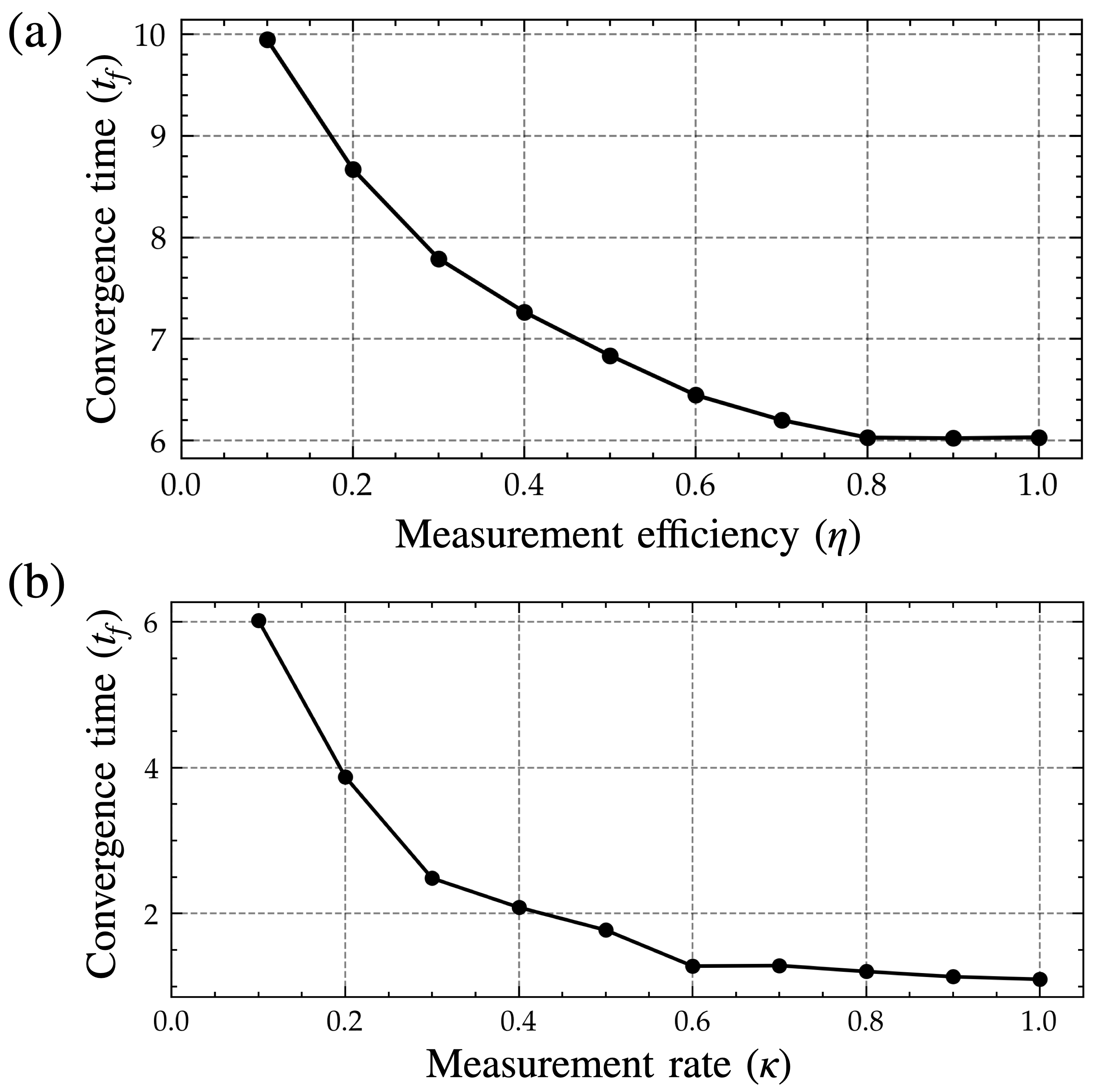}
    \caption{The convergence time $t_f$ as a function of (a) measurement efficiency $\eta$ and (b) measurement rate $\kappa$, showing the fact that $t_f$ depends on the access to information obtained from noisy continuous measurements. }
\label{fig:estimator_show_example1_efficiency_measurement_rate}
\end{figure}

\noindent In the following, we use the intuitive example of a qubit to demonstrate the MBFC protocol, with the Hamiltonian given by, 
\begin{equation}
    \hat H = \frac{\varepsilon \hat \sigma_z}{2} + \frac{\Delta \hat \sigma_x}{2},
\end{equation}
where $\hat \sigma_i, ~ i = (x, y, z)$ are Pauli operators, $\varepsilon$ is the bare energy splitting, and $\Delta$ is the tunneling rate between the two states of the qubit system. We start the state of the physical qubit with excited state occupancy that undergoes continuous measurement of the operator $\hat A=\hat \sigma_z$, and see if the stochastic estimator started with a random state with an initial fidelity of $\sim 0.6$, can lead to a perfect estimate of the state in time. As shown in Fig.~\ref{fig:estimator_show_example1}, the conditional state of the estimator gradually converges with time and perfectly reproduces the real system state. 
Note that this convergence can always be guaranteed regardless of whether the efficiency $\eta$ is ideal or not. In the case of $\eta \neq 1$, the time $t_f$ required to reach convergence becomes slightly longer. This is shown in Fig.~\ref{fig:estimator_show_example1_efficiency_measurement_rate}(a) as a function of $\eta$. On the other hand, for detectors with larger measurement rate $\kappa$, $t_f$ becomes smaller as shown in Fig.~\ref{fig:estimator_show_example1_efficiency_measurement_rate}(b) for $\eta = 1$ as a function of $\kappa$. This behavior of the estimator is understandable, since intuitively the estimator would be able to learn the state faster if it had more accurate information (larger $\eta$) and less noisy measurement data (larger $\kappa$). 

\section{S5. Control laws for symmetric and antisymmetric entangled state preparation}
\noindent We consider the example of two qubits, which starting from random states can be prepared in symmetric and antisymmetric entangled states given by,
\begin{align}
\rho_s = \frac{1}{2}(\psi_{\uparrow\downarrow}+\psi_{\downarrow\uparrow})(\psi_{\uparrow\downarrow}+\psi_{\downarrow\uparrow})^* \\
\rho_a = \frac{1}{2}(\psi_{\uparrow\downarrow}-\psi_{\downarrow\uparrow})(\psi_{\uparrow\downarrow}-\psi_{\downarrow\uparrow})^* ,
\end{align}
where $\psi_{\uparrow \downarrow} = (\uparrow) \otimes (\downarrow)$ and  $\psi_{\downarrow \uparrow } = (\downarrow) \otimes (\uparrow)$ are tensor product states of the individual qubit states in the ground and excited states. We consider the stochastic feedback controls given below~\cite{Mirrahimi2007_qubit_control}.

The quantum filtering equation under feedback with control variables $u_1(t)$ and $u_2(t)$ is given by,
\begin{align}
\nonumber
d\rho(t) = & 
- i u_1(t) [\sigma_y^{(1)}, \rho(t)] dt 
- i u_2(t) [\sigma_y^{(2)}, \rho(t)] dt 
-  \frac{1}{2} [F_z, [F_z, \rho(t)]] dt 
+ \sqrt{\eta} \big\{F_z\rho(t) 
+  \rho(t)F_z \\
-& 2~{\rm Tr}[F_z \rho(t)]\rho(t)\big\} dW_t. 
\end{align}
This can be written as, 
\begin{align}
\nonumber
d\rho(t) = &
- i u_1(t) [\sigma_y^{(1)}, \rho(t)] dt
- i u_2(t) [\sigma_y^{(2)}, \rho(t)] dt 
+ \mathcal{D}[F_z] \rho(t) dt
+ \sqrt{\eta} \mathcal{H}[F_z] \rho(t) dW_t,
\end{align} where $dW_t$ is the Winner noise increment at time $t$. $\sigma_{g}^{i}$, $g \in \{x, y, z\}$ and $i=\{1, 2\}$ are tensored Pauli operators for qubit $i$ and $F_z = \sigma_z^1 + \sigma_z^2$.
This can be rearranged to fit into the the general form of the SME as follows:
\begin{align}
\nonumber
d\rho(t) = &  -i[H, \rho(t)] dt + \mathcal D[A]\rho(t) dt 
+ \mathcal H[A]\rho(t) dW_t,  
\end{align}
where,  
\begin{align}
    H =  ~&  u_1(t) \sigma_y^{(1)} + u_2(t) \sigma_y^{(2)}, \\
    A  =~ &  F_z.
\end{align}
For this, the control laws are given as follows. 
{To stabilize $\rho_a$}, the control laws are:
\begin{enumerate}
    \item $u_1(t)=1-\mathrm{Tr}[{i[\sigma_y^1,\rho]\rho_a}],~
    u_2(t)=1-\mathrm{Tr}[{i[\sigma_y^2,\rho]\rho_a}]$
     if $\mathrm{Tr}[{\rho \rho_a}]\ge\gamma$;
    \item $u_1(t)=1,~u_2(t)=0$ if $\mathrm{Tr}[{\rho\rho_a}]\le\gamma/2$;
    \item If $\rho\in\mathcal{B}_a=\{\rho:\gamma/2<\mathrm{Tr}[{\rho
    \rho_a}]<\gamma\}$,
    then $u_1(t)=1-\mathrm{Tr}[{i[\sigma_y^1,\rho]\rho_a}]$,
    $u_2(t)=1-\mathrm{Tr}[{i[\sigma_y^2,\rho]\rho_a}]$ if $\rho$
    last entered the set $\mathcal{B}_a$ through the boundary
    $\mathrm{Tr}[{\rho\rho_a}]=\gamma$; and $u_1(t)=1,~u_2(t)=0$ otherwise.
\end{enumerate}
Similarly, to stabilize $\rho_s$, the control laws are:
\begin{enumerate}
    \item $u_1(t)=1-\mathrm{Tr}[{i[\sigma_y^1,\rho]\rho_s}],~\\
    u_2(t)=-1-\mathrm{Tr}[{i[\sigma_y^2,\rho]\rho_s}]$
     if $\mathrm{Tr}[{\rho \rho_s}]\ge\gamma$;
    \item $u_1(t)=1,~u_2(t)=0$ if $\mathrm{Tr}[{\rho\rho_s}]\le\gamma/2$;
    \item If $\rho\in\mathcal{B}_s=\{\rho:\gamma/2<\mathrm{Tr}[{\rho
    \rho_s}]<\gamma\}$,
    then take $u_1(t)=1-\mathrm{Tr}[{i[\sigma_y^1,\rho]\rho_s}]$,
    $u_2(t)=-1-\mathrm{Tr}[{i[\sigma_y^2,\rho]\rho_s}]$ if $\rho$
    last entered the set $\mathcal{B}_s$ through the boundary
    $\mathrm{Tr}[{\rho\rho_s}]=\gamma$; and $u_1(t)=1,~u_2(t)=0$ otherwise.
\end{enumerate}

We have demonstrated a convenient representation of the control laws in Fig.~4(a) of the main paper. Note that this feedback control law works only if it is possible to perform real-time tomography of the qubits so that the instantaneous fidelities, $\mathrm{Tr}[\rho\rho_{s/a}]$ with the target symmetric, $\rho_s$ and antisymmetric, $\rho_a$ states can be computed based on which the feedback controls are decided.  

\section{S6. A few more example cases}
\subsection{a. Circuit QED readout and reset}
A fundamentally important problem that is of utmost usefulness in quantum experiments is to reset the state of a unknown quantum state of a qubit. Recently, RL was shown to be possible to discover real-time feedback control for this task~\cite{Reuer2022Oct}. The MBFC protocol proposed in this work enables a single-step reset of the qubit once the estimation is completed.

In circuit quantum electrodynamics (Circuit-QED) platform, the readout is facilitated by coupling with a readout resonator.
We demonstrate the application of the protocol to a Circuit-QED setting, which consists of a coupled resonator-qubit system described by the Jaynes-Cumming (JC) Hamiltonian,
$$
H_{\rm JC} = \frac{\Delta}{2}\sigma_{z} + g(\sigma_+ a + a^\dagger \sigma_-) + \Omega_c (t) (a + a^\dagger),
$$
with the Pauli Z operator $\sigma_z = |g\rangle \langle g| - |e\rangle \langle e|$, $\sigma_- = \sigma_+^\dagger = |g\rangle \langle e|$, and a resonance drive of amplitude $\Omega_c$ on the cavity with the qubit-drive detuning given by $\Delta$. The qubit is coupled to the cavity mode $a$ with coupling strength $g$. In the dispersive limit $\Delta \gg g$, $H_{\rm JC}$ approximates to, 
$$
H_{d} =\frac{1}{2}({\Delta} + \chi ) \sigma_{z} + \chi \sigma _z a^\dagger a + \Omega_c (t) (a + a^\dagger),
$$ which implies a qubit-dependent shift $\chi = {g^2}/{\Delta}$ on the cavity resonance. We choose the following parameters for the numerical simulation of $H_d$:
$g/2\pi = 10$ MHz, $\kappa = 0.2 g$, $\gamma = \gamma_{\phi} = 10^{-4} g$, $|\Omega_c| = 0.173 g$, $\chi = \kappa/2$ and $\Delta/g = 10$. 
\begin{figure}
    \centering
    \includegraphics[width=0.8\linewidth, trim={0cm 0.3cm 0 1cm},clip]{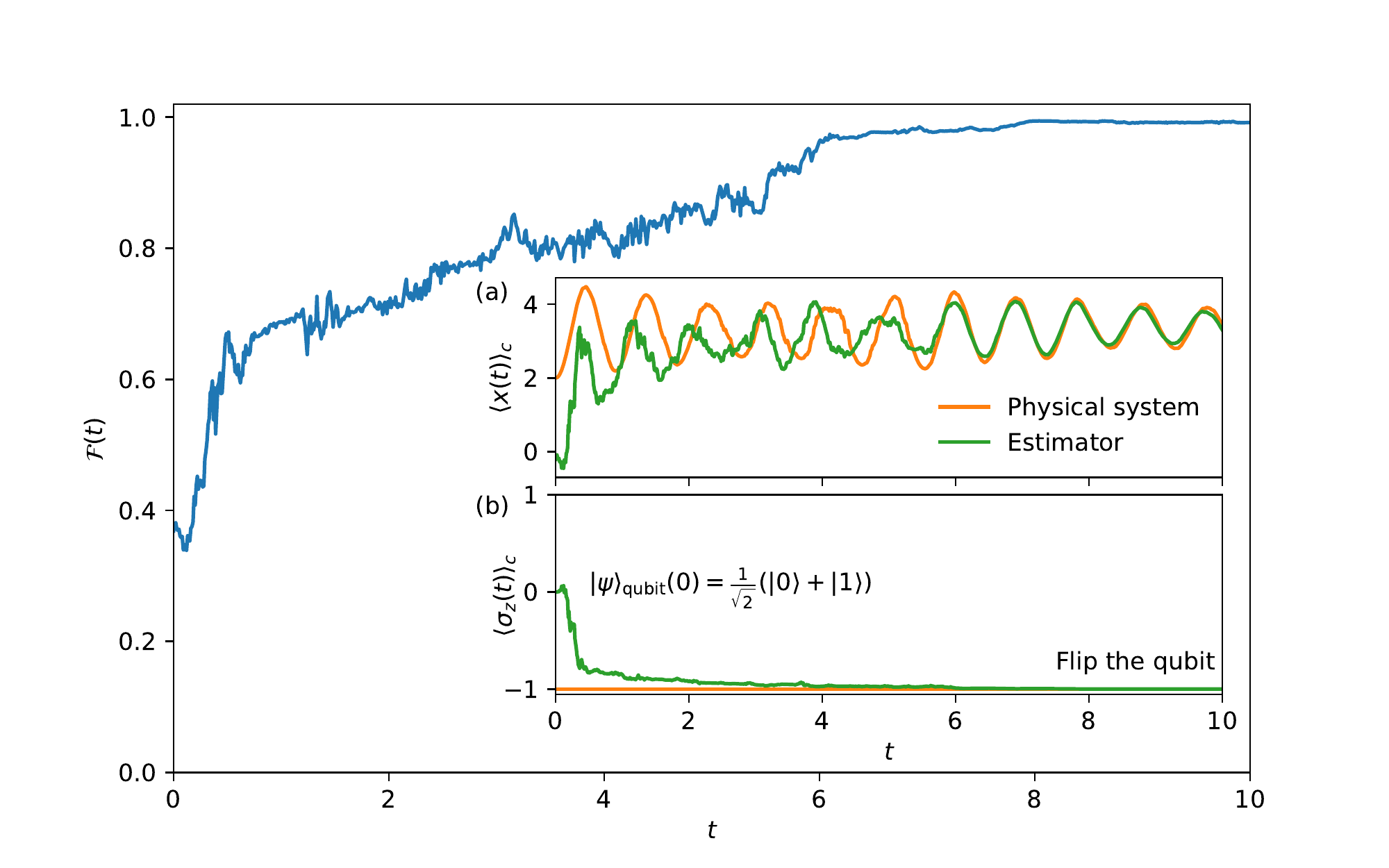}
    \caption{Circuit-QED example. The time evolution of the fidelity between the real and the estimated state is shown. In the inset (a), the $x$-quadrature conditional means $\langle x (t) \rangle_c $ of the real and the estimator system are compared. The corresponding change in $\langle \sigma_z (t) \rangle_c$ of the qubit for the physical system and the estimator are compared in the inset figure (b). The control in this case would be to flip the qubit as $\langle \sigma_z \rangle_c \to -1$.}
    \label{fig:cqed}
\end{figure}

In Fig.~\ref{fig:cqed}, we demonstrate the usability of the protocol to accurately estimate the conditional state of the coupled oscillator-qubit system, with a continuous measurement of the cavity position quadrature operator, $\hat A = x = (a + a^\dagger)/\sqrt{2}$. 
Here, the estimator initial quantum state is considered as a equal superposition of the qubit bases. Once the estimation phase is completed, the qubit can be flipped deterministically using a $\pi$ pulse when $\langle \sigma_z \rangle \to -1$. 
The fidelity of the estimator is plotted with the conditional density matrix of the physical system $\mathcal{F}(t)$, which shows perfect convergence at the end of the estimation step. The inset (a) of the figure compares the conditional expectation value of the position quadrature of the estimator with the physical system, which converges at the end of the estimation phase. The inset (b) of the figure represents the same for the Pauli-$Z$ operator. In this example, the estimator correctly estimated the state of the qubit, which was in the excited state (as $\langle \sigma_z \rangle \sim -1$), that can be subsequently reset by applying a $\pi$ pulse, otherwise for a qubit already in the ground state no reset is applied.

\begin{figure}[!hbt]
    \centering
\includegraphics[width=0.7\linewidth]{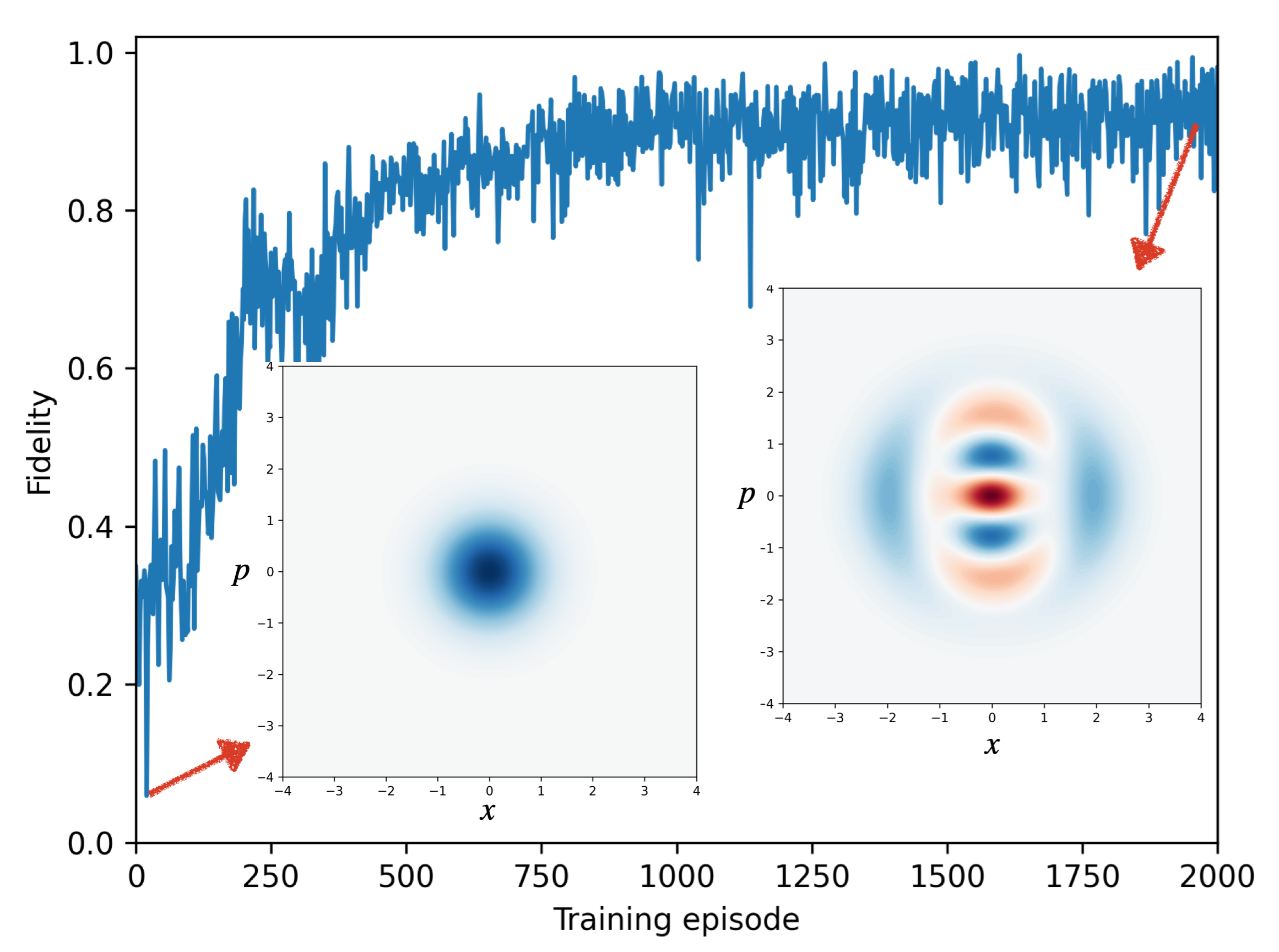}
    \caption{An use-case of the proposed method for non-classical bosonic code preparation using continuous measurement and non-trivial control protocols obtained with RL. The RL-agent is trained to improve the fidelity of a target state (here a binomial 1-3 code is shown) starting from the ground state. The RL can only discover strategies when it gets the instantaneous conditional density matrix as input (observation). The learning curve of the RL is shown. In the insets the wigner function of the initial state and the final state prepared by the RL-agent is demonstrated. The final state has fidelity $> 0.99$.}
    \label{fig:binomial13}
\end{figure}

\subsection{b. State preparation for bosonic error-correcting code}
The proposed MBFC protocol can also be applied efficiently for the generation of bosonic nonclassical states, for example to prepare a binomial 1-3 code. 
Recently, it has been shown that continuous measurement can be utilised to create bosonic non-classical states by applying feedback control determined by RL \cite{Porotti2022Jun}. However, as discussed in that paper, for such a scheme to work one needs to have precise knowledge about the time-continuous changes of the density matrix of the system, which are used as the observation for the RL-agent. Without these, the RL controller cannot discover any good strategy. However time-continuous observation of the system density matrix is a challenge and could be considered nearly as impossible in experiments. For this, our proposed MBFC protocol could provide the necessary solution, as it can discover the precise information through the estimator about the time-continuous density matrix of the to be controlled system, which we briefly describe below.

As described in \cite{Porotti2022Jun}, the problem can be described with the following stochastic master equation, 
\begin{align}
d\rho = - i[F(t), \rho(t)]dt+ \sum_{n} \left[\frac{\kappa_n}{2} \mathcal{D}[P_n]\rho (t) dt + \sqrt{\frac{\kappa_n}{2}} (P_n\rho(t) + \rho (t) P_n - 2\langle P_n \rangle \rho (t))dW_n(t)\right]
\end{align}
where $\mathcal{D}[A]\rho = A\rho A^\dagger - \frac{1}{2} (\rho A^\dagger A + A^\dagger A \rho)$ is the measurement-induced Lindblad superoperator, $\kappa_n$ is the measurement rate of channel $n$, and $dW_n(t)$ is an infinitesimal Wiener process. $F(t)$ is the feedback Hamiltonian of the cavity with a time-dependent drive which can be controlled as 
\begin{equation}
    F(t) = \alpha(t) a^\dagger -  \alpha(t)^* a.
\end{equation}
The measurement operator is given by $P_n = |n\rangle \langle n|$, which represents the projection operator on the $|n\rangle$th fock state of the cavity.
The measurement approach employed in this study is based on the concept of `Multiplexed Photon Number Measurement', as demonstrated in the work of Essig \textit{et al}.~\cite{Essig2021Aug}, which leverages continuous measurements to gather information about the count of photons within a resonator in a time-continuous manner. 

The controls are determined adaptively by a RL agent, where the action is essentially the feedback control $F(t)$, and the reward signal is assigned as the fidelity of the instantaneous state with the target state. The observation-space of the RL-agent is formed by the instantaneous density matrices of the system, for which we found (also shown in ~\cite{Porotti2022Jun}) that the RL-agent must have access to the precise instantaneous matrix elements in order to learn strategies for such a control task. 
We show it with an example, for preparing a bosonic 1-3 code with $\psi_{\rm target} = (|1\rangle + |3\rangle)/\sqrt{2}$ starting from the initial state $|0\rangle$. In Fig.~\ref{fig:binomial13}, the training curve is shown in terms of the improving fidelity between the instantaneous and target states as a function of the training episode. The first inset of the figure on the left hand side shows the Wigner function of the initial state $|0\rangle$. After successful training, the RL agent learns to control the pulses to produce a target state with high fidelity, as shown in the second inset of the figure on the right.    
\begin{figure}[!hbt]
    \centering
    \includegraphics[width=1.0\linewidth, trim={2cm 0 2cm 1cm},clip]{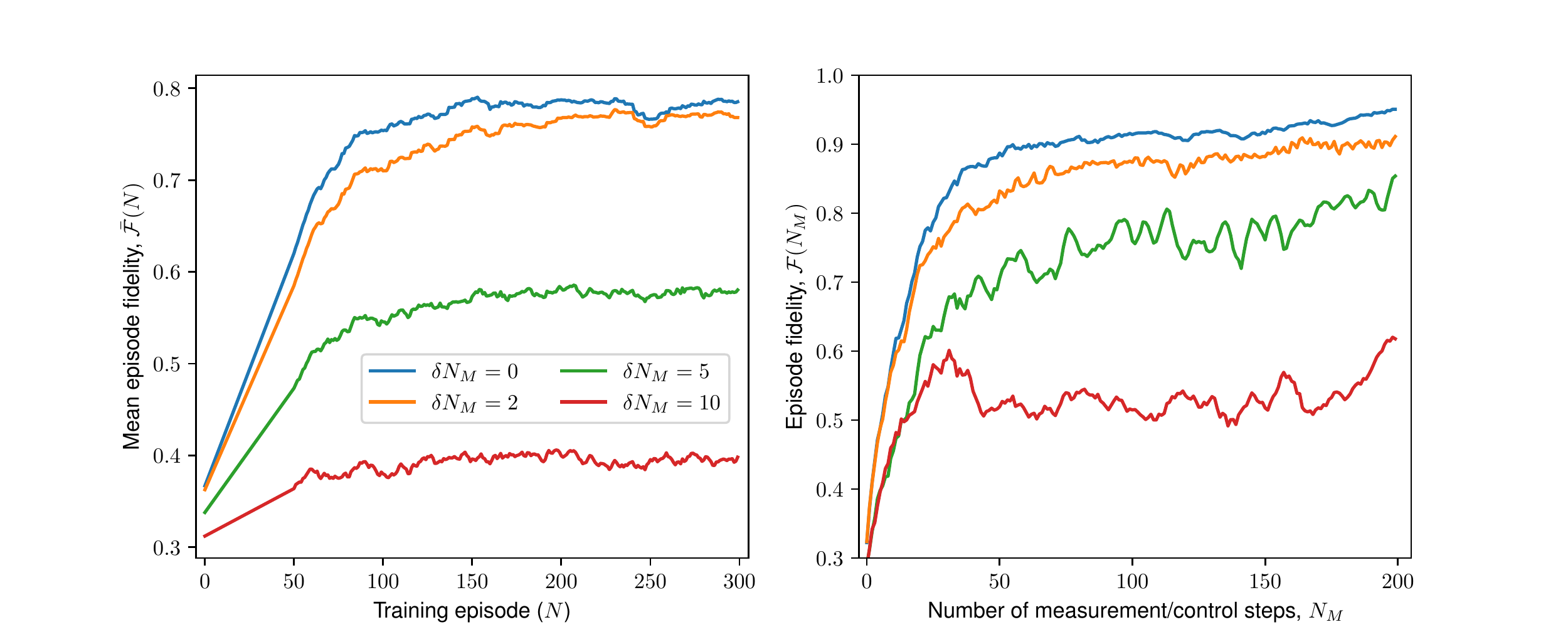}
    \caption{RL to ground state preparation of the double-well potential with continuous measurement control using the protocol. (Left) The training curves of the RL-agent are shown, with the mean episode fidelity, $\bar{\mathcal{F}}(N)$, as a function of the training episode ($N$), for different numbers of measurement step delays, $\delta N_M$, with $\delta N_M = 0$ representing the no-delay case. (Right) The evaluation of the trained RL-agent is shown, with the fidelity, $\mathcal{F}(N_M)$, as a function of $N_M$ (averaged over 100 evaluations) for both no-delay and delayed cases.}
    \label{fig:double_well}
\end{figure}
\subsection{c. Cat state preparation in double-well potential}

In this section, we revisit the problem that was addressed using reinforcement learning (RL) in the recent article ~\cite{Borah2021_double_well}, which focused on finding a non-trivial control to prepare the ground state, i.e.~a cat state in the double-well potential. We briefly outline the problem below.

Considering a quantum double-well potential, the idea is to employ continuous measurement of the operator $\hat{x}^2$, on the basis of which a feedback with a squeezing drive $\lambda(t) \times i(\hat{a}^{\dagger 2} - \hat{a}^2)$ Hamiltonian is applied, with time-varying controls $\lambda(t)$. The measurement operator should not distinguish in which well the particle is, which is why one should not consider $\hat{x}$ as the measurement operator so that the parity of the measurement is preserved. 

In Fig.~\ref{fig:double_well}, the training curves of the RL-agent for the cases considering delays in feedback by a number of measurement steps, $N_M$, with $\delta N_M = 0$ representing no-delay, are depicted on the left. The mean fidelity of the episode, $\bar{\mathcal{F}}(N)$, is shown as a function of the training episode $N$. On the right, the evaluation of the trained RL-agent is displayed as the fidelity, $\mathcal{F}(N_M)$, as a function of $N_M$ (averaged over 100 evaluations) for both no-delay and delayed cases. 

In the original article~\cite{Borah2021_double_well}, the RL-agent was used to discover measurement-based feedback control based on the noisy measurement current, which turned out to be a very difficult task for the model-free RL to manage, requiring more than 1500 episodes of training . By using the MBFC protocol described in this article, it is now possible to use the conditional state of the evolving system as the input (observation) to the RL-agent. For simplicity, we have only considered the diagonal elements of the conditional density matrix. The advantage is remarkable, making it possible to achieve control of the problem within merely 200 episodes.

\section{S7. Comments on the model-based approach} 

\begin{figure}[!hbt]
    \centering
    \includegraphics[width=0.7\linewidth]{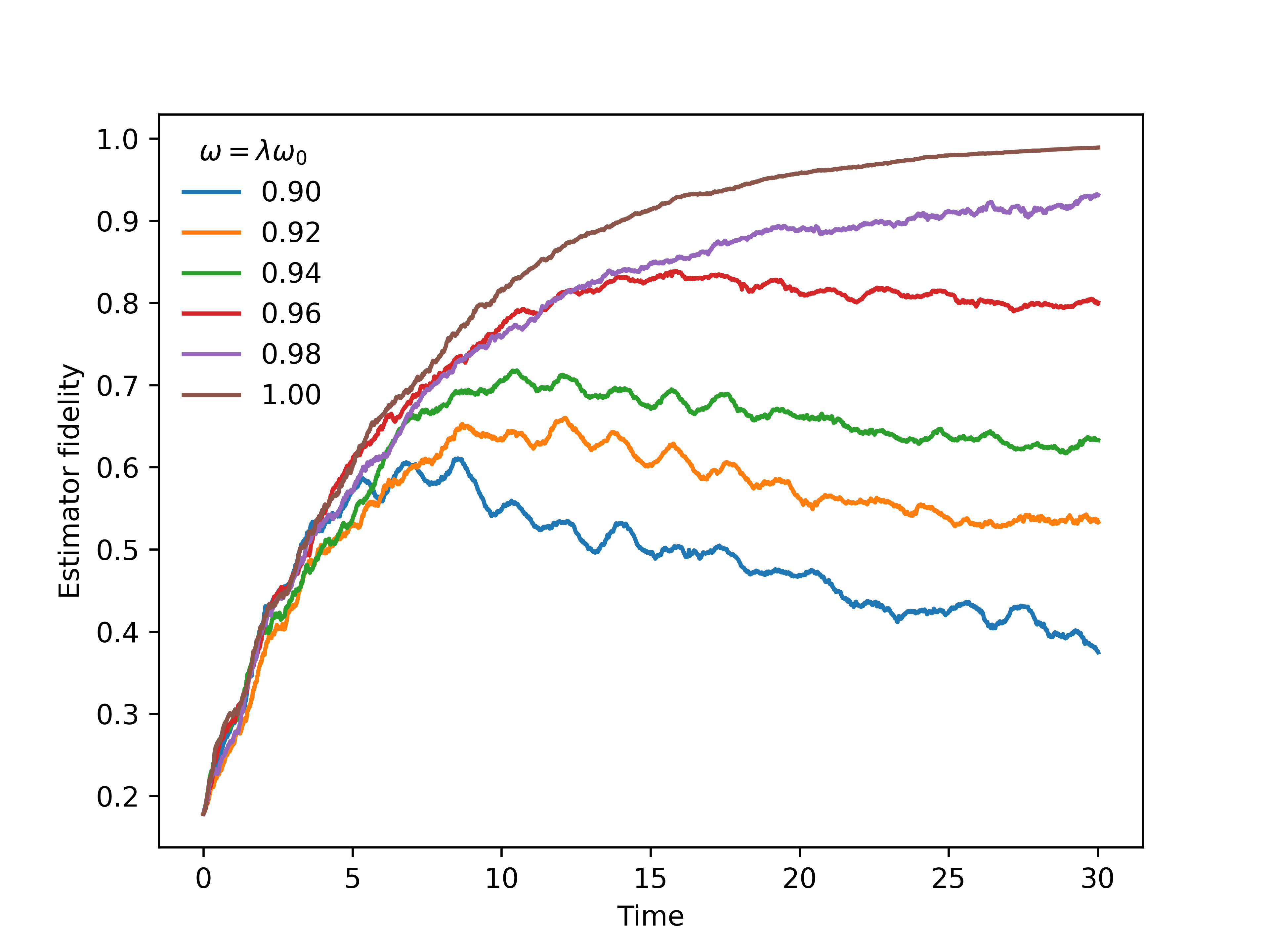}
    \caption{The fidelity of the estimator is demonstrated for a quantum harmonic oscillator Hamiltonian $H=\lambda \omega_0 a^\dagger a$, with $a (a^\dagger)$ being the annihilation (creation) operator of the oscillator and $\omega = \lambda \omega_0$ being the frequency. The legends show the varying values of $\lambda$, which gives the amount of imperfection in knowing the frequency of the oscillator.}
\label{fig:model_based_hamiltonian_dependence}
\end{figure}

The protocol presented in the main paper assumes that we have a very good understanding of the system model so that the protocol works perfectly. It is indeed most suitable for situations with precise and controlled experimental conditions, such as in controlled settings in superconducting circuit quantum electrodynamics systems. We illustrate the effect of not having an accurate estimation of the system Hamiltonian on the estimation phase of the protocol in Fig.~\ref{fig:model_based_hamiltonian_dependence}, for the case of the quantum harmonic oscillator cooling. 

However, as discussed in the main text, as the protocol is expected to benefit from the use of RL-based controllers, the estimation can be improved in realistic experimental situations by combining it with parameter estimation and Hamiltonian learning. There are several methods for such estimation based on measured data, derived from both machine learning and non-machine learning, which could be used before the protocol to estimate the parameters of interest, as summarized in the recent review article on this topic~\cite{Gebhart2022Jul}.


\begin{thebibliography}{47}%
\makeatletter
\providecommand \@ifxundefined [1]{%
 \@ifx{#1\undefined}
}%
\providecommand \@ifnum [1]{%
 \ifnum #1\expandafter \@firstoftwo
 \else \expandafter \@secondoftwo
 \fi
}%
\providecommand \@ifx [1]{%
 \ifx #1\expandafter \@firstoftwo
 \else \expandafter \@secondoftwo
 \fi
}%
\providecommand \natexlab [1]{#1}%
\providecommand \enquote  [1]{``#1''}%
\providecommand \bibnamefont  [1]{#1}%
\providecommand \bibfnamefont [1]{#1}%
\providecommand \citenamefont [1]{#1}%
\providecommand \href@noop [0]{\@secondoftwo}%
\providecommand \href [0]{\begingroup \@sanitize@url \@href}%
\providecommand \@href[1]{\@@startlink{#1}\@@href}%
\providecommand \@@href[1]{\endgroup#1\@@endlink}%
\providecommand \@sanitize@url [0]{\catcode `\\12\catcode `\$12\catcode
  `\&12\catcode `\#12\catcode `\^12\catcode `\_12\catcode `\%12\relax}%
\providecommand \@@startlink[1]{}%
\providecommand \@@endlink[0]{}%
\providecommand \url  [0]{\begingroup\@sanitize@url \@url }%
\providecommand \@url [1]{\endgroup\@href {#1}{\urlprefix }}%
\providecommand \urlprefix  [0]{URL }%
\providecommand \Eprint [0]{\href }%
\providecommand \doibase [0]{https://doi.org/}%
\providecommand \selectlanguage [0]{\@gobble}%
\providecommand \bibinfo  [0]{\@secondoftwo}%
\providecommand \bibfield  [0]{\@secondoftwo}%
\providecommand \translation [1]{[#1]}%
\providecommand \BibitemOpen [0]{}%
\providecommand \bibitemStop [0]{}%
\providecommand \bibitemNoStop [0]{.\EOS\space}%
\providecommand \EOS [0]{\spacefactor3000\relax}%
\providecommand \BibitemShut  [1]{\csname bibitem#1\endcsname}%
\let\auto@bib@innerbib\@empty
\bibitem [{\citenamefont {Wiseman}\ and\ \citenamefont
  {Milburn}(2009)}]{wiseman_milburn_book}%
  \BibitemOpen
  \bibfield  {author} {\bibinfo {author} {\bibfnamefont {H.~M.}\ \bibnamefont
  {Wiseman}}\ and\ \bibinfo {author} {\bibfnamefont {G.~J.}\ \bibnamefont
  {Milburn}},\ }\href {https://doi.org/10.1017/CBO9780511813948} {\emph
  {\bibinfo {title} {{Quantum Measurement and Control}}}}\ (\bibinfo
  {publisher} {Cambridge University Press},\ \bibinfo {address} {Cambridge},\
  \bibinfo {year} {2009})\BibitemShut {NoStop}%
\bibitem [{\citenamefont {Zhang}\ \emph
  {et~al.}(2017{\natexlab{a}})\citenamefont {Zhang}, \citenamefont {Liu},
  \citenamefont {Wu}, \citenamefont {Jacobs},\ and\ \citenamefont
  {Nori}}]{Zhang2017Mar}%
  \BibitemOpen
  \bibfield  {author} {\bibinfo {author} {\bibfnamefont {J.}~\bibnamefont
  {Zhang}}, \bibinfo {author} {\bibfnamefont {Y.-x.}\ \bibnamefont {Liu}},
  \bibinfo {author} {\bibfnamefont {R.-B.}\ \bibnamefont {Wu}}, \bibinfo
  {author} {\bibfnamefont {K.}~\bibnamefont {Jacobs}},\ and\ \bibinfo {author}
  {\bibfnamefont {F.}~\bibnamefont {Nori}},\ }\href
  {https://doi.org/10.1016/j.physrep.2017.02.003} {\bibfield  {journal}
  {\bibinfo  {journal} {Phys. Rep.}\ }\textbf {\bibinfo {volume} {679}},\
  \bibinfo {pages} {1} (\bibinfo {year} {2017}{\natexlab{a}})}\BibitemShut
  {NoStop}%
\bibitem [{\citenamefont {Jacobs}(2014)}]{Jacobs_book}%
  \BibitemOpen
  \bibfield  {author} {\bibinfo {author} {\bibfnamefont {K.}~\bibnamefont
  {Jacobs}},\ }\href {https://doi.org/10.1017/CBO9781139179027} {\emph
  {\bibinfo {title} {{Quantum Measurement Theory and its Applications}}}}\
  (\bibinfo  {publisher} {Cambridge University Press},\ \bibinfo {address}
  {Cambridge, England, UK},\ \bibinfo {year} {2014})\BibitemShut {NoStop}%
\bibitem [{\citenamefont {Doherty}\ \emph {et~al.}(2000)\citenamefont
  {Doherty}, \citenamefont {Habib}, \citenamefont {Jacobs}, \citenamefont
  {Mabuchi},\ and\ \citenamefont {Tan}}]{Doherty2000Jun}%
  \BibitemOpen
  \bibfield  {author} {\bibinfo {author} {\bibfnamefont {A.~C.}\ \bibnamefont
  {Doherty}}, \bibinfo {author} {\bibfnamefont {S.}~\bibnamefont {Habib}},
  \bibinfo {author} {\bibfnamefont {K.}~\bibnamefont {Jacobs}}, \bibinfo
  {author} {\bibfnamefont {H.}~\bibnamefont {Mabuchi}},\ and\ \bibinfo {author}
  {\bibfnamefont {S.~M.}\ \bibnamefont {Tan}},\ }\href
  {https://doi.org/10.1103/PhysRevA.62.012105} {\bibfield  {journal} {\bibinfo
  {journal} {Phys. Rev. A}\ }\textbf {\bibinfo {volume} {62}},\ \bibinfo
  {pages} {012105} (\bibinfo {year} {2000})}\BibitemShut {NoStop}%
\bibitem [{\citenamefont {de~Fouquieres}\ \emph {et~al.}(2011)\citenamefont
  {de~Fouquieres}, \citenamefont {Schirmer}, \citenamefont {Glaser},\ and\
  \citenamefont {Kuprov}}]{deFouquieres2011Oct}%
  \BibitemOpen
  \bibfield  {author} {\bibinfo {author} {\bibfnamefont {P.}~\bibnamefont
  {de~Fouquieres}}, \bibinfo {author} {\bibfnamefont {S.~G.}\ \bibnamefont
  {Schirmer}}, \bibinfo {author} {\bibfnamefont {S.~J.}\ \bibnamefont
  {Glaser}},\ and\ \bibinfo {author} {\bibfnamefont {I.}~\bibnamefont
  {Kuprov}},\ }\href {https://doi.org/10.1016/j.jmr.2011.07.023} {\bibfield
  {journal} {\bibinfo  {journal} {J. Magn. Reson.}\ }\textbf {\bibinfo {volume}
  {212}},\ \bibinfo {pages} {412} (\bibinfo {year} {2011})}\BibitemShut
  {NoStop}%
\bibitem [{\citenamefont {Morzhin}\ and\ \citenamefont
  {Pechen}(2019)}]{Morzhin2019Oct}%
  \BibitemOpen
  \bibfield  {author} {\bibinfo {author} {\bibfnamefont {O.~V.}\ \bibnamefont
  {Morzhin}}\ and\ \bibinfo {author} {\bibfnamefont {A.~N.}\ \bibnamefont
  {Pechen}},\ }\href {https://doi.org/10.1070/rm9835} {\bibfield  {journal}
  {\bibinfo  {journal} {Russ. Math. Surv.}\ }\textbf {\bibinfo {volume} {74}},\
  \bibinfo {pages} {851} (\bibinfo {year} {2019})}\BibitemShut {NoStop}%
\bibitem [{\citenamefont {Koch}\ \emph {et~al.}(2022)\citenamefont {Koch},
  \citenamefont {Boscain}, \citenamefont {Calarco}, \citenamefont {Dirr},
  \citenamefont {Filipp}, \citenamefont {Glaser}, \citenamefont {Kosloff},
  \citenamefont {Montangero}, \citenamefont
  {Schulte-Herbr{\ifmmode\ddot{u}\else\"{u}\fi}ggen}, \citenamefont {Sugny},\
  and\ \citenamefont {Wilhelm}}]{Koch2022Dec}%
  \BibitemOpen
  \bibfield  {author} {\bibinfo {author} {\bibfnamefont {C.~P.}\ \bibnamefont
  {Koch}}, \bibinfo {author} {\bibfnamefont {U.}~\bibnamefont {Boscain}},
  \bibinfo {author} {\bibfnamefont {T.}~\bibnamefont {Calarco}}, \bibinfo
  {author} {\bibfnamefont {G.}~\bibnamefont {Dirr}}, \bibinfo {author}
  {\bibfnamefont {S.}~\bibnamefont {Filipp}}, \bibinfo {author} {\bibfnamefont
  {S.~J.}\ \bibnamefont {Glaser}}, \bibinfo {author} {\bibfnamefont
  {R.}~\bibnamefont {Kosloff}}, \bibinfo {author} {\bibfnamefont
  {S.}~\bibnamefont {Montangero}}, \bibinfo {author} {\bibfnamefont
  {T.}~\bibnamefont {Schulte-Herbr{\ifmmode\ddot{u}\else\"{u}\fi}ggen}},
  \bibinfo {author} {\bibfnamefont {D.}~\bibnamefont {Sugny}},\ and\ \bibinfo
  {author} {\bibfnamefont {F.~K.}\ \bibnamefont {Wilhelm}},\ }\href
  {https://doi.org/10.1140/epjqt/s40507-022-00138-x} {\bibfield  {journal}
  {\bibinfo  {journal} {EPJ Quantum Technol.}\ }\textbf {\bibinfo {volume}
  {9}},\ \bibinfo {pages} {19} (\bibinfo {year} {2022})}\BibitemShut {NoStop}%
\bibitem [{\citenamefont {Sarma}\ \emph {et~al.}(2022)\citenamefont {Sarma},
  \citenamefont {Borah}, \citenamefont {Kani},\ and\ \citenamefont
  {Twamley}}]{Sarma2022Nov}%
  \BibitemOpen
  \bibfield  {author} {\bibinfo {author} {\bibfnamefont {B.}~\bibnamefont
  {Sarma}}, \bibinfo {author} {\bibfnamefont {S.}~\bibnamefont {Borah}},
  \bibinfo {author} {\bibfnamefont {A.}~\bibnamefont {Kani}},\ and\ \bibinfo
  {author} {\bibfnamefont {J.}~\bibnamefont {Twamley}},\ }\href
  {https://doi.org/10.1103/PhysRevResearch.4.L042038} {\bibfield  {journal}
  {\bibinfo  {journal} {Phys. Rev. Res.}\ }\textbf {\bibinfo {volume} {4}},\
  \bibinfo {pages} {L042038} (\bibinfo {year} {2022})}\BibitemShut {NoStop}%
\bibitem [{\citenamefont {Propson}\ \emph {et~al.}(2022)\citenamefont
  {Propson}, \citenamefont {Jackson}, \citenamefont {Koch}, \citenamefont
  {Manchester},\ and\ \citenamefont {Schuster}}]{Propson2022Jan}%
  \BibitemOpen
  \bibfield  {author} {\bibinfo {author} {\bibfnamefont {T.}~\bibnamefont
  {Propson}}, \bibinfo {author} {\bibfnamefont {B.~E.}\ \bibnamefont
  {Jackson}}, \bibinfo {author} {\bibfnamefont {J.}~\bibnamefont {Koch}},
  \bibinfo {author} {\bibfnamefont {Z.}~\bibnamefont {Manchester}},\ and\
  \bibinfo {author} {\bibfnamefont {D.~I.}\ \bibnamefont {Schuster}},\ }\href
  {https://doi.org/10.1103/PhysRevApplied.17.014036} {\bibfield  {journal}
  {\bibinfo  {journal} {Phys. Rev. Appl.}\ }\textbf {\bibinfo {volume} {17}},\
  \bibinfo {pages} {014036} (\bibinfo {year} {2022})}\BibitemShut {NoStop}%
\bibitem [{\citenamefont {Martin}\ \emph {et~al.}(2020)\citenamefont {Martin},
  \citenamefont {Livingston}, \citenamefont {Hacohen-Gourgy}, \citenamefont
  {Wiseman},\ and\ \citenamefont {Siddiqi}}]{Martin2020Oct}%
  \BibitemOpen
  \bibfield  {author} {\bibinfo {author} {\bibfnamefont {L.~S.}\ \bibnamefont
  {Martin}}, \bibinfo {author} {\bibfnamefont {W.~P.}\ \bibnamefont
  {Livingston}}, \bibinfo {author} {\bibfnamefont {S.}~\bibnamefont
  {Hacohen-Gourgy}}, \bibinfo {author} {\bibfnamefont {H.~M.}\ \bibnamefont
  {Wiseman}},\ and\ \bibinfo {author} {\bibfnamefont {I.}~\bibnamefont
  {Siddiqi}},\ }\href {https://doi.org/10.1038/s41567-020-0939-0} {\bibfield
  {journal} {\bibinfo  {journal} {Nat. Phys.}\ }\textbf {\bibinfo {volume}
  {16}},\ \bibinfo {pages} {1046} (\bibinfo {year} {2020})}\BibitemShut
  {NoStop}%
\bibitem [{\citenamefont {Kuang}\ \emph {et~al.}(2021)\citenamefont {Kuang},
  \citenamefont {Li}, \citenamefont {Liu}, \citenamefont {Sun},\ and\
  \citenamefont {Cong}}]{Kuang2021Aug}%
  \BibitemOpen
  \bibfield  {author} {\bibinfo {author} {\bibfnamefont {S.}~\bibnamefont
  {Kuang}}, \bibinfo {author} {\bibfnamefont {G.}~\bibnamefont {Li}}, \bibinfo
  {author} {\bibfnamefont {Y.}~\bibnamefont {Liu}}, \bibinfo {author}
  {\bibfnamefont {X.}~\bibnamefont {Sun}},\ and\ \bibinfo {author}
  {\bibfnamefont {S.}~\bibnamefont {Cong}},\ }\href
  {https://doi.org/10.1109/TCYB.2021.3090676} {\bibfield  {journal} {\bibinfo
  {journal} {IEEE Trans. Cybern.}\ ,\ \bibinfo {pages} {1}} (\bibinfo {year}
  {2021})}\BibitemShut {NoStop}%
\bibitem [{\citenamefont {Rossi}\ \emph {et~al.}(2018)\citenamefont {Rossi},
  \citenamefont {Mason}, \citenamefont {Chen}, \citenamefont {Tsaturyan},\ and\
  \citenamefont {Schliesser}}]{Rossi2018Nov}%
  \BibitemOpen
  \bibfield  {author} {\bibinfo {author} {\bibfnamefont {M.}~\bibnamefont
  {Rossi}}, \bibinfo {author} {\bibfnamefont {D.}~\bibnamefont {Mason}},
  \bibinfo {author} {\bibfnamefont {J.}~\bibnamefont {Chen}}, \bibinfo {author}
  {\bibfnamefont {Y.}~\bibnamefont {Tsaturyan}},\ and\ \bibinfo {author}
  {\bibfnamefont {A.}~\bibnamefont {Schliesser}},\ }\href
  {https://doi.org/10.1038/s41586-018-0643-8} {\bibfield  {journal} {\bibinfo
  {journal} {Nature}\ }\textbf {\bibinfo {volume} {563}},\ \bibinfo {pages}
  {53} (\bibinfo {year} {2018})}\BibitemShut {NoStop}%
\bibitem [{\citenamefont {Vijay}\ \emph {et~al.}(2012)\citenamefont {Vijay},
  \citenamefont {Macklin}, \citenamefont {Slichter}, \citenamefont {Weber},
  \citenamefont {Murch}, \citenamefont {Naik}, \citenamefont {Korotkov},\ and\
  \citenamefont {Siddiqi}}]{Vijay2012Oct}%
  \BibitemOpen
  \bibfield  {author} {\bibinfo {author} {\bibfnamefont {R.}~\bibnamefont
  {Vijay}}, \bibinfo {author} {\bibfnamefont {C.}~\bibnamefont {Macklin}},
  \bibinfo {author} {\bibfnamefont {D.~H.}\ \bibnamefont {Slichter}}, \bibinfo
  {author} {\bibfnamefont {S.~J.}\ \bibnamefont {Weber}}, \bibinfo {author}
  {\bibfnamefont {K.~W.}\ \bibnamefont {Murch}}, \bibinfo {author}
  {\bibfnamefont {R.}~\bibnamefont {Naik}}, \bibinfo {author} {\bibfnamefont
  {A.~N.}\ \bibnamefont {Korotkov}},\ and\ \bibinfo {author} {\bibfnamefont
  {I.}~\bibnamefont {Siddiqi}},\ }\href {https://doi.org/10.1038/nature11505}
  {\bibfield  {journal} {\bibinfo  {journal} {Nature}\ }\textbf {\bibinfo
  {volume} {490}},\ \bibinfo {pages} {77} (\bibinfo {year} {2012})}\BibitemShut
  {NoStop}%
\bibitem [{\citenamefont {Tebbenjohanns}\ \emph {et~al.}(2021)\citenamefont
  {Tebbenjohanns}, \citenamefont {Mattana}, \citenamefont {Rossi},
  \citenamefont {Frimmer},\ and\ \citenamefont
  {Novotny}}]{Tebbenjohanns2021Jul}%
  \BibitemOpen
  \bibfield  {author} {\bibinfo {author} {\bibfnamefont {F.}~\bibnamefont
  {Tebbenjohanns}}, \bibinfo {author} {\bibfnamefont {M.~L.}\ \bibnamefont
  {Mattana}}, \bibinfo {author} {\bibfnamefont {M.}~\bibnamefont {Rossi}},
  \bibinfo {author} {\bibfnamefont {M.}~\bibnamefont {Frimmer}},\ and\ \bibinfo
  {author} {\bibfnamefont {L.}~\bibnamefont {Novotny}},\ }\href
  {https://doi.org/10.1038/s41586-021-03617-w} {\bibfield  {journal} {\bibinfo
  {journal} {Nature}\ }\textbf {\bibinfo {volume} {595}},\ \bibinfo {pages}
  {378} (\bibinfo {year} {2021})}\BibitemShut {NoStop}%
\bibitem [{\citenamefont {Wilson}\ \emph {et~al.}(2015)\citenamefont {Wilson},
  \citenamefont {Sudhir}, \citenamefont {Piro}, \citenamefont {Schilling},
  \citenamefont {Ghadimi},\ and\ \citenamefont {Kippenberg}}]{Wilson2015Aug}%
  \BibitemOpen
  \bibfield  {author} {\bibinfo {author} {\bibfnamefont {D.~J.}\ \bibnamefont
  {Wilson}}, \bibinfo {author} {\bibfnamefont {V.}~\bibnamefont {Sudhir}},
  \bibinfo {author} {\bibfnamefont {N.}~\bibnamefont {Piro}}, \bibinfo {author}
  {\bibfnamefont {R.}~\bibnamefont {Schilling}}, \bibinfo {author}
  {\bibfnamefont {A.}~\bibnamefont {Ghadimi}},\ and\ \bibinfo {author}
  {\bibfnamefont {T.~J.}\ \bibnamefont {Kippenberg}},\ }\href
  {https://doi.org/10.1038/nature14672} {\bibfield  {journal} {\bibinfo
  {journal} {Nature}\ }\textbf {\bibinfo {volume} {524}},\ \bibinfo {pages}
  {325} (\bibinfo {year} {2015})}\BibitemShut {NoStop}%
\bibitem [{\citenamefont {Livingston}\ \emph {et~al.}(2022)\citenamefont
  {Livingston}, \citenamefont {Blok}, \citenamefont {Flurin}, \citenamefont
  {Dressel}, \citenamefont {Jordan},\ and\ \citenamefont
  {Siddiqi}}]{Livingston2022Apr}%
  \BibitemOpen
  \bibfield  {author} {\bibinfo {author} {\bibfnamefont {W.~P.}\ \bibnamefont
  {Livingston}}, \bibinfo {author} {\bibfnamefont {M.~S.}\ \bibnamefont
  {Blok}}, \bibinfo {author} {\bibfnamefont {E.}~\bibnamefont {Flurin}},
  \bibinfo {author} {\bibfnamefont {J.}~\bibnamefont {Dressel}}, \bibinfo
  {author} {\bibfnamefont {A.~N.}\ \bibnamefont {Jordan}},\ and\ \bibinfo
  {author} {\bibfnamefont {I.}~\bibnamefont {Siddiqi}},\ }\href
  {https://doi.org/10.1038/s41467-022-29906-0} {\bibfield  {journal} {\bibinfo
  {journal} {Nat. Commun.}\ }\textbf {\bibinfo {volume} {13}},\ \bibinfo
  {pages} {1} (\bibinfo {year} {2022})}\BibitemShut {NoStop}%
\bibitem [{\citenamefont {Magrini}\ \emph {et~al.}(2021)\citenamefont
  {Magrini}, \citenamefont {Rosenzweig}, \citenamefont {Bach}, \citenamefont
  {Deutschmann-Olek}, \citenamefont {Hofer}, \citenamefont {Hong},
  \citenamefont {Kiesel}, \citenamefont {Kugi},\ and\ \citenamefont
  {Aspelmeyer}}]{Magrini2021Jul}%
  \BibitemOpen
  \bibfield  {author} {\bibinfo {author} {\bibfnamefont {L.}~\bibnamefont
  {Magrini}}, \bibinfo {author} {\bibfnamefont {P.}~\bibnamefont {Rosenzweig}},
  \bibinfo {author} {\bibfnamefont {C.}~\bibnamefont {Bach}}, \bibinfo {author}
  {\bibfnamefont {A.}~\bibnamefont {Deutschmann-Olek}}, \bibinfo {author}
  {\bibfnamefont {S.~G.}\ \bibnamefont {Hofer}}, \bibinfo {author}
  {\bibfnamefont {S.}~\bibnamefont {Hong}}, \bibinfo {author} {\bibfnamefont
  {N.}~\bibnamefont {Kiesel}}, \bibinfo {author} {\bibfnamefont
  {A.}~\bibnamefont {Kugi}},\ and\ \bibinfo {author} {\bibfnamefont
  {M.}~\bibnamefont {Aspelmeyer}},\ }\href
  {https://doi.org/10.1038/s41586-021-03602-3} {\bibfield  {journal} {\bibinfo
  {journal} {Nature}\ }\textbf {\bibinfo {volume} {595}},\ \bibinfo {pages}
  {373} (\bibinfo {year} {2021})}\BibitemShut {NoStop}%
\bibitem [{\citenamefont {Jim\'enez-Mart\'{\i}nez}\ \emph
  {et~al.}(2018)\citenamefont {Jim\'enez-Mart\'{\i}nez}, \citenamefont
  {Ko\l{}ody\ifmmode~\acute{n}\else \'{n}\fi{}ski}, \citenamefont {Troullinou},
  \citenamefont {Lucivero}, \citenamefont {Kong},\ and\ \citenamefont
  {Mitchell}}]{JimenezMartinez2018Jan}%
  \BibitemOpen
  \bibfield  {author} {\bibinfo {author} {\bibfnamefont {R.}~\bibnamefont
  {Jim\'enez-Mart\'{\i}nez}}, \bibinfo {author} {\bibfnamefont
  {J.}~\bibnamefont {Ko\l{}ody\ifmmode~\acute{n}\else \'{n}\fi{}ski}}, \bibinfo
  {author} {\bibfnamefont {C.}~\bibnamefont {Troullinou}}, \bibinfo {author}
  {\bibfnamefont {V.~G.}\ \bibnamefont {Lucivero}}, \bibinfo {author}
  {\bibfnamefont {J.}~\bibnamefont {Kong}},\ and\ \bibinfo {author}
  {\bibfnamefont {M.~W.}\ \bibnamefont {Mitchell}},\ }\href
  {https://doi.org/10.1103/PhysRevLett.120.040503} {\bibfield  {journal}
  {\bibinfo  {journal} {Phys. Rev. Lett.}\ }\textbf {\bibinfo {volume} {120}},\
  \bibinfo {pages} {040503} (\bibinfo {year} {2018})}\BibitemShut {NoStop}%
\bibitem [{\citenamefont {F{\"{o}}sel}\ \emph {et~al.}(2018)\citenamefont
  {F{\"{o}}sel}, \citenamefont {Tighineanu}, \citenamefont {Weiss},\ and\
  \citenamefont {Marquardt}}]{Marquardt2018_drl_qec}%
  \BibitemOpen
  \bibfield  {author} {\bibinfo {author} {\bibfnamefont {T.}~\bibnamefont
  {F{\"{o}}sel}}, \bibinfo {author} {\bibfnamefont {P.}~\bibnamefont
  {Tighineanu}}, \bibinfo {author} {\bibfnamefont {T.}~\bibnamefont {Weiss}},\
  and\ \bibinfo {author} {\bibfnamefont {F.}~\bibnamefont {Marquardt}},\
  }\href@noop {} {\bibfield  {journal} {\bibinfo  {journal} {Phys. Rev. X}\
  }\textbf {\bibinfo {volume} {8}} (\bibinfo {year} {2018})}\BibitemShut
  {NoStop}%
\bibitem [{\citenamefont {Bukov}\ \emph {et~al.}(2018)\citenamefont {Bukov},
  \citenamefont {Day}, \citenamefont {Sels}, \citenamefont {Weinberg},
  \citenamefont {Polkovnikov},\ and\ \citenamefont {Mehta}}]{Bukov2018Sep}%
  \BibitemOpen
  \bibfield  {author} {\bibinfo {author} {\bibfnamefont {M.}~\bibnamefont
  {Bukov}}, \bibinfo {author} {\bibfnamefont {A.~G.~R.}\ \bibnamefont {Day}},
  \bibinfo {author} {\bibfnamefont {D.}~\bibnamefont {Sels}}, \bibinfo {author}
  {\bibfnamefont {P.}~\bibnamefont {Weinberg}}, \bibinfo {author}
  {\bibfnamefont {A.}~\bibnamefont {Polkovnikov}},\ and\ \bibinfo {author}
  {\bibfnamefont {P.}~\bibnamefont {Mehta}},\ }\href
  {https://doi.org/10.1103/PhysRevX.8.031086} {\bibfield  {journal} {\bibinfo
  {journal} {Phys. Rev. X}\ }\textbf {\bibinfo {volume} {8}},\ \bibinfo {pages}
  {031086} (\bibinfo {year} {2018})}\BibitemShut {NoStop}%
\bibitem [{\citenamefont {Wang}\ \emph {et~al.}(2020)\citenamefont {Wang},
  \citenamefont {Ashida},\ and\ \citenamefont
  {Ueda}}]{Ueda2020_quantum_cartpole_drl}%
  \BibitemOpen
  \bibfield  {author} {\bibinfo {author} {\bibfnamefont {Z.~T.}\ \bibnamefont
  {Wang}}, \bibinfo {author} {\bibfnamefont {Y.}~\bibnamefont {Ashida}},\ and\
  \bibinfo {author} {\bibfnamefont {M.}~\bibnamefont {Ueda}},\ }\href@noop {}
  {\bibfield  {journal} {\bibinfo  {journal} {Phys. Rev. Lett.}\ }\textbf
  {\bibinfo {volume} {125}},\ \bibinfo {pages} {100401} (\bibinfo {year}
  {2020})}\BibitemShut {NoStop}%
\bibitem [{\citenamefont {Niu}\ \emph {et~al.}(2019)\citenamefont {Niu},
  \citenamefont {Boixo}, \citenamefont {Smelyanskiy},\ and\ \citenamefont
  {Neven}}]{Niu2019Apr}%
  \BibitemOpen
  \bibfield  {author} {\bibinfo {author} {\bibfnamefont {M.~Y.}\ \bibnamefont
  {Niu}}, \bibinfo {author} {\bibfnamefont {S.}~\bibnamefont {Boixo}}, \bibinfo
  {author} {\bibfnamefont {V.~N.}\ \bibnamefont {Smelyanskiy}},\ and\ \bibinfo
  {author} {\bibfnamefont {H.}~\bibnamefont {Neven}},\ }\href@noop {}
  {\bibfield  {journal} {\bibinfo  {journal} {npj Quantum Inf.}\ }\textbf
  {\bibinfo {volume} {5}},\ \bibinfo {pages} {1} (\bibinfo {year}
  {2019})}\BibitemShut {NoStop}%
\bibitem [{\citenamefont {Zhang}\ \emph
  {et~al.}(2019{\natexlab{a}})\citenamefont {Zhang}, \citenamefont {Wei},
  \citenamefont {Asad}, \citenamefont {Yang},\ and\ \citenamefont
  {Wang}}]{Zhang2019Oct}%
  \BibitemOpen
  \bibfield  {author} {\bibinfo {author} {\bibfnamefont {X.-M.}\ \bibnamefont
  {Zhang}}, \bibinfo {author} {\bibfnamefont {Z.}~\bibnamefont {Wei}}, \bibinfo
  {author} {\bibfnamefont {R.}~\bibnamefont {Asad}}, \bibinfo {author}
  {\bibfnamefont {X.-C.}\ \bibnamefont {Yang}},\ and\ \bibinfo {author}
  {\bibfnamefont {X.}~\bibnamefont {Wang}},\ }\href@noop {} {\bibfield
  {journal} {\bibinfo  {journal} {npj Quantum Inf.}\ }\textbf {\bibinfo
  {volume} {5}},\ \bibinfo {pages} {1} (\bibinfo {year}
  {2019}{\natexlab{a}})}\BibitemShut {NoStop}%
\bibitem [{\citenamefont {Borah}\ \emph {et~al.}(2021)\citenamefont {Borah},
  \citenamefont {Sarma}, \citenamefont {Kewming}, \citenamefont {Milburn},\
  and\ \citenamefont {Twamley}}]{Borah2021_double_well}%
  \BibitemOpen
  \bibfield  {author} {\bibinfo {author} {\bibfnamefont {S.}~\bibnamefont
  {Borah}}, \bibinfo {author} {\bibfnamefont {B.}~\bibnamefont {Sarma}},
  \bibinfo {author} {\bibfnamefont {M.}~\bibnamefont {Kewming}}, \bibinfo
  {author} {\bibfnamefont {G.~J.}\ \bibnamefont {Milburn}},\ and\ \bibinfo
  {author} {\bibfnamefont {J.}~\bibnamefont {Twamley}},\ }\href
  {https://doi.org/10.1103/PhysRevLett.127.190403} {\bibfield  {journal}
  {\bibinfo  {journal} {Phys. Rev. Lett.}\ }\textbf {\bibinfo {volume} {127}},\
  \bibinfo {pages} {190403} (\bibinfo {year} {2021})}\BibitemShut {NoStop}%
\bibitem [{\citenamefont {Porotti}\ \emph {et~al.}(2019)\citenamefont
  {Porotti}, \citenamefont {Tamascelli}, \citenamefont {Restelli},\ and\
  \citenamefont {Prati}}]{Prati2019_drl_stirap}%
  \BibitemOpen
  \bibfield  {author} {\bibinfo {author} {\bibfnamefont {R.}~\bibnamefont
  {Porotti}}, \bibinfo {author} {\bibfnamefont {D.}~\bibnamefont {Tamascelli}},
  \bibinfo {author} {\bibfnamefont {M.}~\bibnamefont {Restelli}},\ and\
  \bibinfo {author} {\bibfnamefont {E.}~\bibnamefont {Prati}},\ }\href@noop {}
  {\bibfield  {journal} {\bibinfo  {journal} {{Commun. Phys.}}\ }\textbf
  {\bibinfo {volume} {{2}}} (\bibinfo {year} {{2019}})}\BibitemShut {NoStop}%
\bibitem [{\citenamefont {Paparelle}\ \emph {et~al.}(2020)\citenamefont
  {Paparelle}, \citenamefont {Moro},\ and\ \citenamefont
  {Prati}}]{Paparelle2020}%
  \BibitemOpen
  \bibfield  {author} {\bibinfo {author} {\bibfnamefont {I.}~\bibnamefont
  {Paparelle}}, \bibinfo {author} {\bibfnamefont {L.}~\bibnamefont {Moro}},\
  and\ \bibinfo {author} {\bibfnamefont {E.}~\bibnamefont {Prati}},\
  }\href@noop {} {\bibfield  {journal} {\bibinfo  {journal} {Phys. Lett. A}\
  }\textbf {\bibinfo {volume} {384}},\ \bibinfo {pages} {126266} (\bibinfo
  {year} {2020})}\BibitemShut {NoStop}%
\bibitem [{\citenamefont {Porotti}\ \emph {et~al.}(2022)\citenamefont
  {Porotti}, \citenamefont {Essig}, \citenamefont {Huard},\ and\ \citenamefont
  {Marquardt}}]{Porotti2022Jun}%
  \BibitemOpen
  \bibfield  {author} {\bibinfo {author} {\bibfnamefont {R.}~\bibnamefont
  {Porotti}}, \bibinfo {author} {\bibfnamefont {A.}~\bibnamefont {Essig}},
  \bibinfo {author} {\bibfnamefont {B.}~\bibnamefont {Huard}},\ and\ \bibinfo
  {author} {\bibfnamefont {F.}~\bibnamefont {Marquardt}},\ }\href
  {https://doi.org/10.22331/q-2022-06-28-747} {\bibfield  {journal} {\bibinfo
  {journal} {Quantum}\ }\textbf {\bibinfo {volume} {6}},\ \bibinfo {pages}
  {747} (\bibinfo {year} {2022})},\ \Eprint
  {https://arxiv.org/abs/2107.08816v3} {2107.08816v3} \BibitemShut {NoStop}%
\bibitem [{\citenamefont {Zhang}\ \emph
  {et~al.}(2019{\natexlab{b}})\citenamefont {Zhang}, \citenamefont {Wei},
  \citenamefont {Asad}, \citenamefont {Yang},\ and\ \citenamefont
  {Wang}}]{Wang2019_drl_state_prepare}%
  \BibitemOpen
  \bibfield  {author} {\bibinfo {author} {\bibfnamefont {X.-M.}\ \bibnamefont
  {Zhang}}, \bibinfo {author} {\bibfnamefont {Z.}~\bibnamefont {Wei}}, \bibinfo
  {author} {\bibfnamefont {R.}~\bibnamefont {Asad}}, \bibinfo {author}
  {\bibfnamefont {X.-C.}\ \bibnamefont {Yang}},\ and\ \bibinfo {author}
  {\bibfnamefont {X.}~\bibnamefont {Wang}},\ }\href@noop {} {\bibfield
  {journal} {\bibinfo  {journal} {npj Quantum Inf.}\ }\textbf {\bibinfo
  {volume} {{5}}} (\bibinfo {year} {{2019}}{\natexlab{b}})}\BibitemShut
  {NoStop}%
\bibitem [{\citenamefont {Mackeprang}\ \emph {et~al.}(2020)\citenamefont
  {Mackeprang}, \citenamefont {{Rao Dasari}},\ and\ \citenamefont
  {Wrachtrup}}]{Wrachtrup2020_drl_state_engineering}%
  \BibitemOpen
  \bibfield  {author} {\bibinfo {author} {\bibfnamefont {J.}~\bibnamefont
  {Mackeprang}}, \bibinfo {author} {\bibfnamefont {D.~B.}\ \bibnamefont {{Rao
  Dasari}}},\ and\ \bibinfo {author} {\bibfnamefont {J.}~\bibnamefont
  {Wrachtrup}},\ }\href@noop {} {\bibfield  {journal} {\bibinfo  {journal}
  {QuantumMachine Intell.}\ }\textbf {\bibinfo {volume} {2}},\ \bibinfo {pages}
  {1} (\bibinfo {year} {2020})}\BibitemShut {NoStop}%
\bibitem [{\citenamefont {Haug}\ \emph {et~al.}(2020)\citenamefont {Haug},
  \citenamefont {Mok}, \citenamefont {You}, \citenamefont {Zhang},
  \citenamefont {{Eng Png}},\ and\ \citenamefont
  {Kwek}}]{Haug2020_drl_state_preparation_classification}%
  \BibitemOpen
  \bibfield  {author} {\bibinfo {author} {\bibfnamefont {T.}~\bibnamefont
  {Haug}}, \bibinfo {author} {\bibfnamefont {W.-K.}\ \bibnamefont {Mok}},
  \bibinfo {author} {\bibfnamefont {J.-B.}\ \bibnamefont {You}}, \bibinfo
  {author} {\bibfnamefont {W.}~\bibnamefont {Zhang}}, \bibinfo {author}
  {\bibfnamefont {C.}~\bibnamefont {{Eng Png}}},\ and\ \bibinfo {author}
  {\bibfnamefont {L.-C.}\ \bibnamefont {Kwek}},\ }\href@noop {} {\bibfield
  {journal} {\bibinfo  {journal} {Mach. Learn. Sci. Technol.}\ }\textbf
  {\bibinfo {volume} {2}} (\bibinfo {year} {2020})}\BibitemShut {NoStop}%
\bibitem [{\citenamefont {Nautrup}\ \emph {et~al.}(2019)\citenamefont
  {Nautrup}, \citenamefont {Delfosse}, \citenamefont {Dunjko}, \citenamefont
  {Briegel},\ and\ \citenamefont {Friis}}]{Nautrup2019Dec}%
  \BibitemOpen
  \bibfield  {author} {\bibinfo {author} {\bibfnamefont {H.~P.}\ \bibnamefont
  {Nautrup}}, \bibinfo {author} {\bibfnamefont {N.}~\bibnamefont {Delfosse}},
  \bibinfo {author} {\bibfnamefont {V.}~\bibnamefont {Dunjko}}, \bibinfo
  {author} {\bibfnamefont {H.~J.}\ \bibnamefont {Briegel}},\ and\ \bibinfo
  {author} {\bibfnamefont {N.}~\bibnamefont {Friis}},\ }\href@noop {}
  {\bibfield  {journal} {\bibinfo  {journal} {Quantum}\ }\textbf {\bibinfo
  {volume} {3}},\ \bibinfo {pages} {215} (\bibinfo {year} {2019})}\BibitemShut
  {NoStop}%
\bibitem [{\citenamefont {Reuer}\ \emph {et~al.}(2022)\citenamefont {Reuer},
  \citenamefont {Landgraf}, \citenamefont
  {F{\ifmmode\ddot{o}\else\"{o}\fi}sel}, \citenamefont {O'Sullivan},
  \citenamefont {Beltr{\ifmmode\acute{a}\else\'{a}\fi}n}, \citenamefont {Akin},
  \citenamefont {Norris}, \citenamefont {Remm}, \citenamefont {Kerschbaum},
  \citenamefont {Besse}, \citenamefont {Marquardt}, \citenamefont {Wallraff},\
  and\ \citenamefont {Eichler}}]{Reuer2022Oct}%
  \BibitemOpen
  \bibfield  {author} {\bibinfo {author} {\bibfnamefont {K.}~\bibnamefont
  {Reuer}}, \bibinfo {author} {\bibfnamefont {J.}~\bibnamefont {Landgraf}},
  \bibinfo {author} {\bibfnamefont {T.}~\bibnamefont
  {F{\ifmmode\ddot{o}\else\"{o}\fi}sel}}, \bibinfo {author} {\bibfnamefont
  {J.}~\bibnamefont {O'Sullivan}}, \bibinfo {author} {\bibfnamefont
  {L.}~\bibnamefont {Beltr{\ifmmode\acute{a}\else\'{a}\fi}n}}, \bibinfo
  {author} {\bibfnamefont {A.}~\bibnamefont {Akin}}, \bibinfo {author}
  {\bibfnamefont {G.~J.}\ \bibnamefont {Norris}}, \bibinfo {author}
  {\bibfnamefont {A.}~\bibnamefont {Remm}}, \bibinfo {author} {\bibfnamefont
  {M.}~\bibnamefont {Kerschbaum}}, \bibinfo {author} {\bibfnamefont {J.-C.}\
  \bibnamefont {Besse}}, \bibinfo {author} {\bibfnamefont {F.}~\bibnamefont
  {Marquardt}}, \bibinfo {author} {\bibfnamefont {A.}~\bibnamefont
  {Wallraff}},\ and\ \bibinfo {author} {\bibfnamefont {C.}~\bibnamefont
  {Eichler}},\ }\bibfield  {journal} {\bibinfo  {journal} {arXiv}\ }\href
  {https://doi.org/10.48550/arXiv.2210.16715} {10.48550/arXiv.2210.16715}
  (\bibinfo {year} {2022}),\ \Eprint {https://arxiv.org/abs/2210.16715}
  {2210.16715} \BibitemShut {NoStop}%
\bibitem [{\citenamefont {Sivak}\ \emph {et~al.}(2023)\citenamefont {Sivak},
  \citenamefont {Eickbusch}, \citenamefont {Royer}, \citenamefont {Singh},
  \citenamefont {Tsioutsios}, \citenamefont {Ganjam}, \citenamefont {Miano},
  \citenamefont {Brock}, \citenamefont {Ding}, \citenamefont {Frunzio},
  \citenamefont {Girvin}, \citenamefont {Schoelkopf},\ and\ \citenamefont
  {Devoret}}]{Sivak2022Nov}%
  \BibitemOpen
  \bibfield  {author} {\bibinfo {author} {\bibfnamefont {V.~V.}\ \bibnamefont
  {Sivak}}, \bibinfo {author} {\bibfnamefont {A.}~\bibnamefont {Eickbusch}},
  \bibinfo {author} {\bibfnamefont {B.}~\bibnamefont {Royer}}, \bibinfo
  {author} {\bibfnamefont {S.}~\bibnamefont {Singh}}, \bibinfo {author}
  {\bibfnamefont {I.}~\bibnamefont {Tsioutsios}}, \bibinfo {author}
  {\bibfnamefont {S.}~\bibnamefont {Ganjam}}, \bibinfo {author} {\bibfnamefont
  {A.}~\bibnamefont {Miano}}, \bibinfo {author} {\bibfnamefont {B.~L.}\
  \bibnamefont {Brock}}, \bibinfo {author} {\bibfnamefont {A.~Z.}\ \bibnamefont
  {Ding}}, \bibinfo {author} {\bibfnamefont {L.}~\bibnamefont {Frunzio}},
  \bibinfo {author} {\bibfnamefont {S.~M.}\ \bibnamefont {Girvin}}, \bibinfo
  {author} {\bibfnamefont {R.~J.}\ \bibnamefont {Schoelkopf}},\ and\ \bibinfo
  {author} {\bibfnamefont {M.~H.}\ \bibnamefont {Devoret}},\ }\href
  {https://doi.org/10.1038/s41586-023-05782-6} {\bibfield  {journal} {\bibinfo
  {journal} {Nature}\ }\textbf {\bibinfo {volume} {616}},\ \bibinfo {pages}
  {50} (\bibinfo {year} {2023})}\BibitemShut {NoStop}%
\bibitem [{\citenamefont {Hacohen-Gourgy}\ and\ \citenamefont
  {Martin}(2020)}]{Hacohen-Gourgy2020Jan}%
  \BibitemOpen
  \bibfield  {author} {\bibinfo {author} {\bibfnamefont {S.}~\bibnamefont
  {Hacohen-Gourgy}}\ and\ \bibinfo {author} {\bibfnamefont {L.~S.}\
  \bibnamefont {Martin}},\ }\href
  {https://doi.org/10.1080/23746149.2020.1813626} {\bibfield  {journal}
  {\bibinfo  {journal} {Advances in Physics: X}\ }\textbf {\bibinfo {volume}
  {5}},\ \bibinfo {pages} {1813626} (\bibinfo {year} {2020})}\BibitemShut
  {NoStop}%
\bibitem [{\citenamefont {Zhang}\ \emph
  {et~al.}(2017{\natexlab{b}})\citenamefont {Zhang}, \citenamefont {Liu},
  \citenamefont {Wu}, \citenamefont {Jacobs},\ and\ \citenamefont
  {Nori}}]{Zhang2017}%
  \BibitemOpen
  \bibfield  {author} {\bibinfo {author} {\bibfnamefont {J.}~\bibnamefont
  {Zhang}}, \bibinfo {author} {\bibfnamefont {Y.-x.}\ \bibnamefont {Liu}},
  \bibinfo {author} {\bibfnamefont {R.-B.}\ \bibnamefont {Wu}}, \bibinfo
  {author} {\bibfnamefont {K.}~\bibnamefont {Jacobs}},\ and\ \bibinfo {author}
  {\bibfnamefont {F.}~\bibnamefont {Nori}},\ }\href@noop {} {\bibfield
  {journal} {\bibinfo  {journal} {Phys. Rep.}\ }\textbf {\bibinfo {volume}
  {679}},\ \bibinfo {pages} {1} (\bibinfo {year}
  {2017}{\natexlab{b}})}\BibitemShut {NoStop}%
\bibitem [{\citenamefont {Di{\ifmmode\acute{o}\else\'{o}\fi}si}\ \emph
  {et~al.}(2006)\citenamefont {Di{\ifmmode\acute{o}\else\'{o}\fi}si},
  \citenamefont {Konrad}, \citenamefont {Scherer},\ and\ \citenamefont
  {Audretsch}}]{Diosi2006_coupled_ito}%
  \BibitemOpen
  \bibfield  {author} {\bibinfo {author} {\bibfnamefont {L.}~\bibnamefont
  {Di{\ifmmode\acute{o}\else\'{o}\fi}si}}, \bibinfo {author} {\bibfnamefont
  {T.}~\bibnamefont {Konrad}}, \bibinfo {author} {\bibfnamefont
  {A.}~\bibnamefont {Scherer}},\ and\ \bibinfo {author} {\bibfnamefont
  {J.}~\bibnamefont {Audretsch}},\ }\href
  {https://doi.org/10.1088/0305-4470/39/40/l01} {\bibfield  {journal} {\bibinfo
   {journal} {J. Phys. A: Math. Gen.}\ }\textbf {\bibinfo {volume} {39}},\
  \bibinfo {pages} {L575} (\bibinfo {year} {2006})}\BibitemShut {NoStop}%
\bibitem [{\citenamefont {Sutton}\ and\ \citenamefont
  {Barto}(2018)}]{Sutton_drl_book}%
  \BibitemOpen
  \bibfield  {author} {\bibinfo {author} {\bibfnamefont {R.~S.}\ \bibnamefont
  {Sutton}}\ and\ \bibinfo {author} {\bibfnamefont {A.~G.}\ \bibnamefont
  {Barto}},\ }\href@noop {} {\emph {\bibinfo {title} {{Reinforcement Learning:
  An Introduction}}}},\ \bibinfo {edition} {2nd}\ ed.\ (\bibinfo  {publisher}
  {The MIT Press},\ \bibinfo {year} {2018})\BibitemShut {NoStop}%
\bibitem [{\citenamefont {Schulman}\ \emph
  {et~al.}(2017{\natexlab{a}})\citenamefont {Schulman}, \citenamefont {Wolski},
  \citenamefont {Dhariwal}, \citenamefont {Radford},\ and\ \citenamefont
  {Klimov}}]{ppo_paper}%
  \BibitemOpen
  \bibfield  {author} {\bibinfo {author} {\bibfnamefont {J.}~\bibnamefont
  {Schulman}}, \bibinfo {author} {\bibfnamefont {F.}~\bibnamefont {Wolski}},
  \bibinfo {author} {\bibfnamefont {P.}~\bibnamefont {Dhariwal}}, \bibinfo
  {author} {\bibfnamefont {A.}~\bibnamefont {Radford}},\ and\ \bibinfo {author}
  {\bibfnamefont {O.}~\bibnamefont {Klimov}},\ }\href@noop {} {\bibinfo {title}
  {Proximal policy optimization algorithms}} (\bibinfo {year}
  {2017}{\natexlab{a}}),\ \Eprint {https://arxiv.org/abs/1707.06347}
  {arXiv:1707.06347 [cs.LG]} \BibitemShut {NoStop}%
\bibitem [{\citenamefont {Schulman}\ \emph
  {et~al.}(2017{\natexlab{b}})\citenamefont {Schulman}, \citenamefont {Levine},
  \citenamefont {Moritz}, \citenamefont {Jordan},\ and\ \citenamefont
  {Abbeel}}]{trpo_paper}%
  \BibitemOpen
  \bibfield  {author} {\bibinfo {author} {\bibfnamefont {J.}~\bibnamefont
  {Schulman}}, \bibinfo {author} {\bibfnamefont {S.}~\bibnamefont {Levine}},
  \bibinfo {author} {\bibfnamefont {P.}~\bibnamefont {Moritz}}, \bibinfo
  {author} {\bibfnamefont {M.~I.}\ \bibnamefont {Jordan}},\ and\ \bibinfo
  {author} {\bibfnamefont {P.}~\bibnamefont {Abbeel}},\ }\href@noop {}
  {\bibinfo {title} {Trust region policy optimization}} (\bibinfo {year}
  {2017}{\natexlab{b}}),\ \Eprint {https://arxiv.org/abs/1502.05477}
  {arXiv:1502.05477 [cs.LG]} \BibitemShut {NoStop}%
\bibitem [{\citenamefont {Mirrahimi}\ and\ \citenamefont
  {Van~Handel}(2007)}]{Mirrahimi2007_qubit_control}%
  \BibitemOpen
  \bibfield  {author} {\bibinfo {author} {\bibfnamefont {M.}~\bibnamefont
  {Mirrahimi}}\ and\ \bibinfo {author} {\bibfnamefont {R.}~\bibnamefont
  {Van~Handel}},\ }\href {https://epubs.siam.org/doi/abs/10.1137/050644793}
  {\bibfield  {journal} {\bibinfo  {journal} {SIAM J. Control Optim.}\ }
  (\bibinfo {year} {2007})}\BibitemShut {NoStop}%
\bibitem [{\citenamefont {Gebhart}\ \emph
  {et~al.}(2023{\natexlab{a}})\citenamefont {Gebhart}, \citenamefont
  {Santagati}, \citenamefont {Gentile}, \citenamefont {Gauger}, \citenamefont
  {Craig}, \citenamefont {Ares}, \citenamefont {Banchi}, \citenamefont
  {Marquardt}, \citenamefont {Pezz{\ifmmode\grave{e}\else\`{e}\fi}},\ and\
  \citenamefont {Bonato}}]{Gebhart2023Mar}%
  \BibitemOpen
  \bibfield  {author} {\bibinfo {author} {\bibfnamefont {V.}~\bibnamefont
  {Gebhart}}, \bibinfo {author} {\bibfnamefont {R.}~\bibnamefont {Santagati}},
  \bibinfo {author} {\bibfnamefont {A.~A.}\ \bibnamefont {Gentile}}, \bibinfo
  {author} {\bibfnamefont {E.~M.}\ \bibnamefont {Gauger}}, \bibinfo {author}
  {\bibfnamefont {D.}~\bibnamefont {Craig}}, \bibinfo {author} {\bibfnamefont
  {N.}~\bibnamefont {Ares}}, \bibinfo {author} {\bibfnamefont {L.}~\bibnamefont
  {Banchi}}, \bibinfo {author} {\bibfnamefont {F.}~\bibnamefont {Marquardt}},
  \bibinfo {author} {\bibfnamefont {L.}~\bibnamefont
  {Pezz{\ifmmode\grave{e}\else\`{e}\fi}}},\ and\ \bibinfo {author}
  {\bibfnamefont {C.}~\bibnamefont {Bonato}},\ }\href
  {https://doi.org/10.1038/s42254-022-00552-1} {\bibfield  {journal} {\bibinfo
  {journal} {Nat. Rev. Phys.}\ }\textbf {\bibinfo {volume} {5}},\ \bibinfo
  {pages} {141} (\bibinfo {year} {2023}{\natexlab{a}})}\BibitemShut {NoStop}%
\bibitem [{\citenamefont {Nolan}\ \emph {et~al.}(2021)\citenamefont {Nolan},
  \citenamefont {Smerzi},\ and\ \citenamefont
  {Pezz{\ifmmode\grave{e}\else\`{e}\fi}}}]{Nolan2021Dec}%
  \BibitemOpen
  \bibfield  {author} {\bibinfo {author} {\bibfnamefont {S.}~\bibnamefont
  {Nolan}}, \bibinfo {author} {\bibfnamefont {A.}~\bibnamefont {Smerzi}},\ and\
  \bibinfo {author} {\bibfnamefont {L.}~\bibnamefont
  {Pezz{\ifmmode\grave{e}\else\`{e}\fi}}},\ }\href
  {https://doi.org/10.1038/s41534-021-00497-w} {\bibfield  {journal} {\bibinfo
  {journal} {npj Quantum Inf.}\ }\textbf {\bibinfo {volume} {7}},\ \bibinfo
  {pages} {1} (\bibinfo {year} {2021})}\BibitemShut {NoStop}%
\bibitem [{\citenamefont {Xiao}\ \emph {et~al.}(2022)\citenamefont {Xiao},
  \citenamefont {Fan},\ and\ \citenamefont {Zeng}}]{Xiao2022Jan}%
  \BibitemOpen
  \bibfield  {author} {\bibinfo {author} {\bibfnamefont {T.}~\bibnamefont
  {Xiao}}, \bibinfo {author} {\bibfnamefont {J.}~\bibnamefont {Fan}},\ and\
  \bibinfo {author} {\bibfnamefont {G.}~\bibnamefont {Zeng}},\ }\href
  {https://doi.org/10.1038/s41534-021-00513-z} {\bibfield  {journal} {\bibinfo
  {journal} {npj Quantum Inf.}\ }\textbf {\bibinfo {volume} {8}},\ \bibinfo
  {pages} {1} (\bibinfo {year} {2022})}\BibitemShut {NoStop}%
\bibitem [{\citenamefont {Xu}\ \emph {et~al.}(2019)\citenamefont {Xu},
  \citenamefont {Li}, \citenamefont {Liu}, \citenamefont {Wang}, \citenamefont
  {Yuan},\ and\ \citenamefont {Wang}}]{Xu2019Oct}%
  \BibitemOpen
  \bibfield  {author} {\bibinfo {author} {\bibfnamefont {H.}~\bibnamefont
  {Xu}}, \bibinfo {author} {\bibfnamefont {J.}~\bibnamefont {Li}}, \bibinfo
  {author} {\bibfnamefont {L.}~\bibnamefont {Liu}}, \bibinfo {author}
  {\bibfnamefont {Y.}~\bibnamefont {Wang}}, \bibinfo {author} {\bibfnamefont
  {H.}~\bibnamefont {Yuan}},\ and\ \bibinfo {author} {\bibfnamefont
  {X.}~\bibnamefont {Wang}},\ }\href
  {https://doi.org/10.1038/s41534-019-0198-z} {\bibfield  {journal} {\bibinfo
  {journal} {npj Quantum Inf.}\ }\textbf {\bibinfo {volume} {5}},\ \bibinfo
  {pages} {1} (\bibinfo {year} {2019})}\BibitemShut {NoStop}%
\bibitem [{\citenamefont {Goodfellow}\ \emph {et~al.}(2016)\citenamefont
  {Goodfellow}, \citenamefont {Bengio}, \citenamefont {Courville},\ and\
  \citenamefont {Bengio}}]{Goodfellow_ml_book}%
  \BibitemOpen
  \bibfield  {author} {\bibinfo {author} {\bibfnamefont {I.}~\bibnamefont
  {Goodfellow}}, \bibinfo {author} {\bibfnamefont {Y.}~\bibnamefont {Bengio}},
  \bibinfo {author} {\bibfnamefont {A.}~\bibnamefont {Courville}},\ and\
  \bibinfo {author} {\bibfnamefont {Y.}~\bibnamefont {Bengio}},\ }\href@noop {}
  {\emph {\bibinfo {title} {Deep learning}}},\ Vol.~\bibinfo {volume} {1}\
  (\bibinfo  {publisher} {MIT press Cambridge},\ \bibinfo {year}
  {2016})\BibitemShut {NoStop}%
\bibitem [{\citenamefont {Essig}\ \emph {et~al.}(2021)\citenamefont {Essig},
  \citenamefont {Ficheux}, \citenamefont {Peronnin}, \citenamefont {Cottet},
  \citenamefont {Lescanne}, \citenamefont {Sarlette}, \citenamefont {Rouchon},
  \citenamefont {Leghtas},\ and\ \citenamefont {Huard}}]{Essig2021Aug}%
  \BibitemOpen
  \bibfield  {author} {\bibinfo {author} {\bibfnamefont {A.}~\bibnamefont
  {Essig}}, \bibinfo {author} {\bibfnamefont {Q.}~\bibnamefont {Ficheux}},
  \bibinfo {author} {\bibfnamefont {T.}~\bibnamefont {Peronnin}}, \bibinfo
  {author} {\bibfnamefont {N.}~\bibnamefont {Cottet}}, \bibinfo {author}
  {\bibfnamefont {R.}~\bibnamefont {Lescanne}}, \bibinfo {author}
  {\bibfnamefont {A.}~\bibnamefont {Sarlette}}, \bibinfo {author}
  {\bibfnamefont {P.}~\bibnamefont {Rouchon}}, \bibinfo {author} {\bibfnamefont
  {Z.}~\bibnamefont {Leghtas}},\ and\ \bibinfo {author} {\bibfnamefont
  {B.}~\bibnamefont {Huard}},\ }\href
  {https://doi.org/10.1103/PhysRevX.11.031045} {\bibfield  {journal} {\bibinfo
  {journal} {Phys. Rev. X}\ }\textbf {\bibinfo {volume} {11}},\ \bibinfo
  {pages} {031045} (\bibinfo {year} {2021})}\BibitemShut {NoStop}%
\bibitem [{\citenamefont {Gebhart}\ \emph
  {et~al.}(2023{\natexlab{b}})\citenamefont {Gebhart}, \citenamefont
  {Santagati}, \citenamefont {Gentile}, \citenamefont {Gauger}, \citenamefont
  {Craig}, \citenamefont {Ares}, \citenamefont {Banchi}, \citenamefont
  {Marquardt}, \citenamefont {Pezz{\ifmmode\grave{e}\else\`{e}\fi}},\ and\
  \citenamefont {Bonato}}]{Gebhart2022Jul}%
  \BibitemOpen
  \bibfield  {author} {\bibinfo {author} {\bibfnamefont {V.}~\bibnamefont
  {Gebhart}}, \bibinfo {author} {\bibfnamefont {R.}~\bibnamefont {Santagati}},
  \bibinfo {author} {\bibfnamefont {A.~A.}\ \bibnamefont {Gentile}}, \bibinfo
  {author} {\bibfnamefont {E.~M.}\ \bibnamefont {Gauger}}, \bibinfo {author}
  {\bibfnamefont {D.}~\bibnamefont {Craig}}, \bibinfo {author} {\bibfnamefont
  {N.}~\bibnamefont {Ares}}, \bibinfo {author} {\bibfnamefont {L.}~\bibnamefont
  {Banchi}}, \bibinfo {author} {\bibfnamefont {F.}~\bibnamefont {Marquardt}},
  \bibinfo {author} {\bibfnamefont {L.}~\bibnamefont
  {Pezz{\ifmmode\grave{e}\else\`{e}\fi}}},\ and\ \bibinfo {author}
  {\bibfnamefont {C.}~\bibnamefont {Bonato}},\ }\href
  {https://doi.org/10.1038/s42254-022-00552-1} {\bibfield  {journal} {\bibinfo
  {journal} {Nat. Rev. Phys.}\ }\textbf {\bibinfo {volume} {5}},\ \bibinfo
  {pages} {141} (\bibinfo {year} {2023}{\natexlab{b}})}\BibitemShut {NoStop}%
\end{thebibliography}

\end{document}